\documentclass[12pt,a4paper]{article}
\ifdefined\pdfminorversion\pdfminorversion=7\fi

\usepackage[T1]{fontenc}
\usepackage{lmodern}
\usepackage[utf8]{inputenc}
\usepackage[english]{babel}
\usepackage[margin=1in]{geometry}
\usepackage{setspace}
\usepackage{amsmath,amssymb,bm}
\usepackage{textcomp}
\usepackage{authblk}
\usepackage{graphicx}
\usepackage{caption}
\usepackage{subfigure}
\usepackage{booktabs}
\usepackage{float}
\usepackage{placeins}
\usepackage{needspace}
\usepackage[authoryear,round]{natbib}
\usepackage{etoolbox}
\usepackage{microtype}
\usepackage{xurl}
\usepackage{hyperref}

\graphicspath{{./frames/}}
\hypersetup{colorlinks=true,linkcolor=blue,citecolor=blue,urlcolor=blue}
\AtBeginEnvironment{table}{\small\setlength{\tabcolsep}{4pt}}

\title{Robust Functional Ward\textquotesingle s Linkages with Applications in EEG Data Clustering}

\author[1]{Tianbo Chen}
\author[2]{Wenlin Dai}
\author[1]{Hanbing Zhu\thanks{Corresponding author: Hanbing Zhu. Email: \href{mailto:zhuhbecnu@163.com}{\nolinkurl{zhuhbecnu@163.com}}.}}
\affil[1]{School of Big Data and Statistics, Anhui University, Hefei, China}
\affil[2]{Institute of Statistics and Big Data, Renmin University of China, Beijing, China}
\date{}

\begin{document}
\maketitle

\begin{abstract}
This paper proposes two robust linkage criteria, termed functional Ward-type linkages, for hierarchical clustering of grouped functional data, where each clustering unit is represented by a set of curves. Ordinary Ward\textquotesingle s linkage defines the cluster dissimilarity between two clusters as the increase in the sum of squared errors (SSE) upon merging, which can be interpreted graphically as an increase in within-cluster dispersion. Motivated by this principle, the proposed linkages measure the weighted increase in band width required to construct a common band containing the central representatives of two candidate clusters. To reduce the influence of outliers, the central curves are selected separately within each cluster by leveraging the magnitude-shape outlyingness measure and the modified band depth, respectively. Simulations and two EEG data analyses demonstrate the robustness
		and competitive performance of the proposed methods under a range
		of contamination settings.
\end{abstract}

\noindent\textbf{Keywords:} EEG data analysis; Functional data clustering; Functional Ward-type linkage; Magnitude-shape outlyingness; Robustness.

\section{Introduction}
	
	{ Functional data clustering identifies homogeneous groups when observations are curves or functions rather than finite-dimensional vectors. It has been widely used in energy, environmental, biomedical, and time-series applications; examples include water-demand modeling \citep{cheifetz2017modeling} and shape-based demand-response clustering \citep{teeraratkul2017shape}. General reviews are provided by Wang et al. \citeyearpar{wang2016functional} and Zhang and Parnell \citeyearpar{zhang2023review}. Time-series and spectral-density clustering form an important application area because a spectral density estimate can be treated as a functional observation. For EEG data, Eu\'an et al. \citeyearpar{euan2018hierarchical} clustered channels using distances between spectral densities, while Maadooliat et al. \citeyearpar{maadooliat2018nonparametric} and Chen et al. \citeyearpar{chen2020collective} developed collective spectral-estimation and clustering procedures. The problem studied here is broader: we consider grouped functional data in which each observational unit is represented by a set of curves, as may arise from repeated measurements on the same subject, recording, channel, experimental unit, or time period. Some curves within a set may be affected by measurement errors, artifacts, abnormal variation, or other contamination. The objective is therefore to cluster sets of curves according to their representative central functional behavior, rather than to cluster individual curves or vectorized functional summaries.
		
		Existing functional clustering methods may be broadly organized into procedures based on the original curves, fitted models, dimension reduction, distances or linkages, and model-based formulations \citep{liao2005clustering,jacques2014functional}. Related reviews for functional and time-series clustering are given by Montero and Vilar \citeyearpar{montero2015tsclust}, Aghabozorgi et al. \citeyearpar{aghabozorgi2015time}, Chamroukhi and Nguyen \citeyearpar{chamroukhi2019model}, and Cheam and Fredette \citeyearpar{cheam2020importance}. Most of these approaches treat an individual curve or time series as the clustering unit. Robustness is especially important for functional data because an atypical curve may involve a global magnitude shift, a localized shape change, an unusual covariance pattern, or a short-lived artifact. A common robustification strategy is trimming. Garcia-Escudero and Gordaliza \citeyearpar{garcia2005proposal} proposed trimmed $k$-means for spline coefficients, Rivera-Garc\'ia et al. \citeyearpar{rivera2019robust} incorporated trimming into model-based functional clustering, and related approaches were developed by Cuesta-Albertos et al. \citeyearpar{cuesta1997trimmed}, Cuesta-Albertos and Fraiman \citeyearpar{cuesta2007impartial}, Delaigle and Hall \citeyearpar{delaigle2010defining}, and Kondo et al. \citeyearpar{kondo2016rskc}. Trimming can improve stability by excluding observations least compatible with the fitted structure, but its performance depends on the trimming level and clustering loss. Trimmed observations may remain unassigned, and contaminated curves that resemble the center of another cluster may not be recognized as globally outlying.
		
		Robust procedures without explicit global trimming have also been proposed. Chen et al. \citeyearpar{chen2021clustering} represented a cluster using either the functional median or the 50\% most central curves defined through functional depth \citep{lopez2009concept}. A single median is robust but does not reflect the variation among representative curves, whereas fixing the retained proportion at 50\% may be restrictive. D\textquotesingle Urso et al. \citeyearpar{d2015time,d2016garch} developed robust medoid- and GARCH-based time-series clustering methods whose performance can depend on model specification. The present paper is positioned primarily within robust functional data clustering, with a specific contribution to hierarchical linkage construction for grouped functional observations. Ward\textquotesingle s minimum-variance method \citep{ward1963hierarchical} merges the pair of clusters that produces the smallest increase in within-cluster sum of squared errors; its Hilbert-space analogue uses functional means and squared Hilbert-space norms. Related properties and extensions are discussed by Milligan \citeyearpar{milligan1979ultrametric}, Strauss and Von Maltitz \citeyearpar{strauss2017generalising}, Sharma et al. \citeyearpar{sharma2019comparative}, and Großwendt et al. \citeyearpar{grosswendt2019analysis}. We adopt a hierarchical framework because it yields a multi-resolution dendrogram and requires only a cluster-level dissimilarity. In our setting, dispersion is described geometrically through the width of a functional band. A band constructed from all curves, however, is determined by pointwise extrema and can be strongly affected by a few atypical curves.
		
		To address this limitation, we propose two robust functional Ward-type linkage criteria based on central subsets identified by the magnitude--shape (MS) plot \citep{dai2018multivariate,dai2019directional} and modified band depth (MBD) \citep{lopez2009concept,sun2011functional}. Within each candidate cluster, curves are ranked from the center outward, a central subset is selected, and the cluster is represented by the band generated by these curves. For each candidate pair, the two central subsets are selected separately, and the linkage measures the weighted increase in band width required to construct a common band containing both subsets. The construction is trimming-inspired because only a selected central proportion contributes directly to the linkage calculation. Unlike conventional trimming, however, noncentral curves are not permanently removed or left unassigned; they remain members of their current clusters, and centrality is recomputed after clusters are merged. The central band retains information about variation among representative curves while limiting the effect of peripheral observations. In addition, the MS and MBD rankings provide information on whether atypical curves primarily reflect magnitude, shape, or depth-based deviations. We compare the proposed linkages with directly aligned set-level hierarchical methods using the same initial groups and agglomerative framework. Generic robust clustering procedures applied to individual curves or finite-dimensional representations are retained only as external pipeline-level benchmarks because they differ in clustering unit, algorithm, and robustness mechanism.
		
		The main contributions are fourfold. First, we formulate a grouped functional clustering problem in which each clustering unit contains multiple curves and may include contamination. Second, we introduce two robust nonnegative hierarchical linkage criteria based on MS- and MBD-central subsets. Third, we compare them with directly aligned set-level linkages and examine sensitivity to initial-cluster impurity, initial-cluster size, and the central proportion. Fourth, we illustrate the framework using EEG spectral-density clustering. In these applications, repeated segments from the same recording form an observational set, and diagnostic or experimental labels are not used to construct the initial clusters or determine the agglomerative merging sequence. Section~\ref{EP} introduces the methodology, Section~\ref{SM} presents simulations and competing methods, Section~\ref{APP} gives the EEG applications, and Section~\ref{CO} concludes the paper.}
	
	\section{Methodology}\label{EP}
	
	\subsection{Functional Ward\textquotesingle s Linkage}\label{CWD}
	Ordinary Ward\textquotesingle s linkage defines the squared distance between two clusters, denoted as $\boldsymbol{C_1}$ and $\boldsymbol{C}_2$, as the increase in the SSE when they are merged into a new cluster $\boldsymbol{C}_1\cup \boldsymbol{C}_2$:
	$$
	D^2(\boldsymbol{C}_1, \boldsymbol{C}_2)= {\rm SSE}(\boldsymbol{C}_1\cup \boldsymbol{C}_2) - {\rm SSE}(\boldsymbol{C}_1) - {\rm SSE}(\boldsymbol{C}_2).
	$$
	Intuitively, SSE represents the diameter, often interpreted as within-cluster dispersion:
	$$
	{\rm dia}(\boldsymbol{C}_i) =\text{SSE}(\boldsymbol{C}_i)  = \sum_{j=1}^{|\boldsymbol{C}_i|} (y_{ij}-\bar{y}_i)^{\top} (y_{ij}-\bar{y}_i) = |\boldsymbol{C}_i|\bar{d}_{\boldsymbol{C}_i},\quad i=1,2,
	$$
	where ${\rm dia}(\cdot)$ denotes the within-cluster dispersion, $y_{ij}$ represents the $j$-th sample in cluster $\boldsymbol{C}_i$, $\bar{y_i}$ is the centroid of cluster  $\boldsymbol{C}_i$, $\bar{d}_{\boldsymbol{C}_i}$ is the average squared distance to the centroid, and $|\cdot|$ is the number of elements in a cluster. Then, 
	\begin{equation} \label{eq1}
		D^2(\boldsymbol{C}_1, \boldsymbol{C}_2) = |\boldsymbol{C}_1\cup \boldsymbol{C}_2|\bar{d}_{\boldsymbol{C}_1\cup \boldsymbol{C}_2} - |\boldsymbol{C}_1|\bar{d}_{\boldsymbol{C}_1} -|\boldsymbol{C}_2|\bar{d}_{\boldsymbol{C}_2}.
	\end{equation}
	This implies that merging more homogeneous clusters leads to a smaller increase in the cluster within-cluster dispersion.

\begin{figure}[htbp]
		\begin{center}
			\subfigure[]{\raisebox{-0.3cm}{\includegraphics[width=0.6\textwidth]{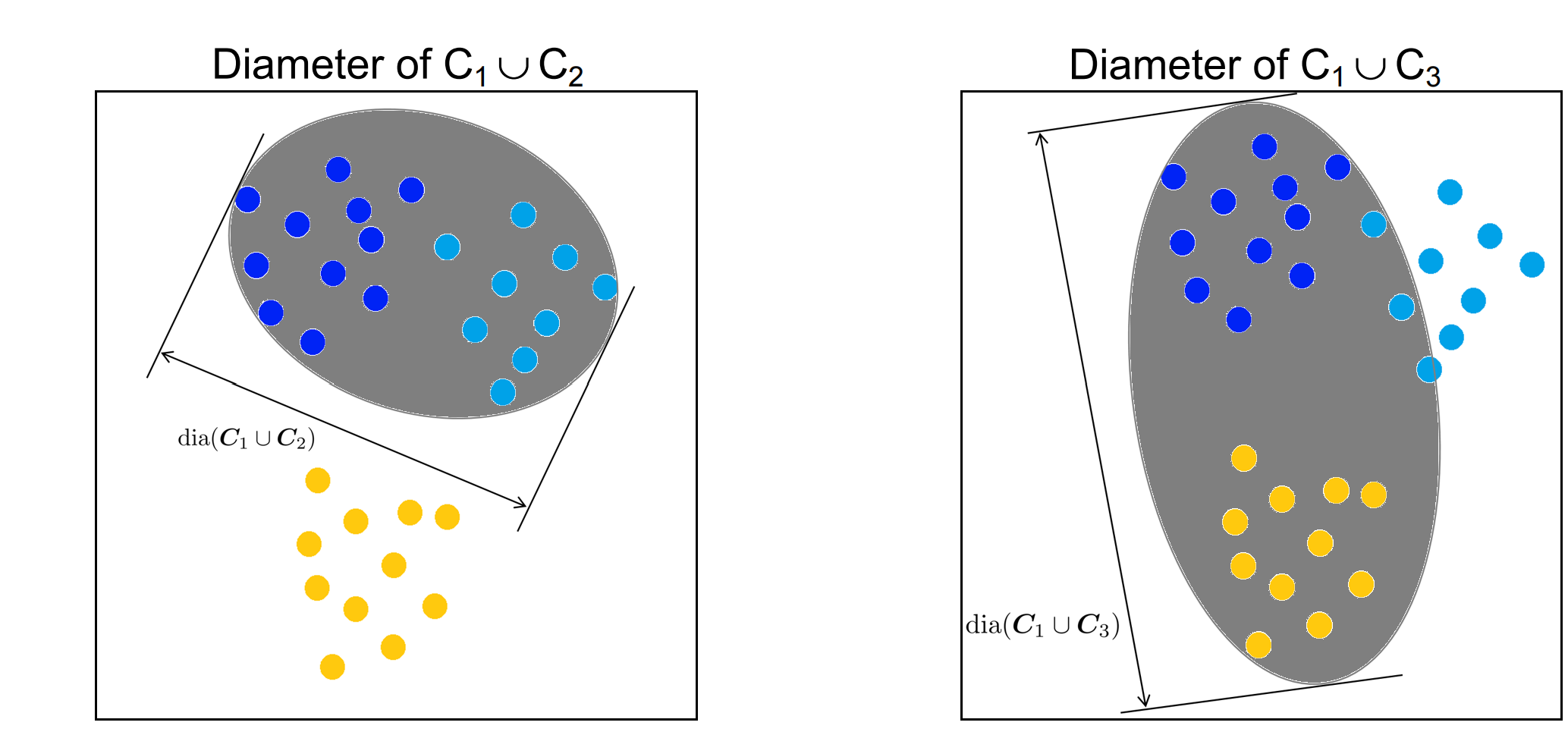}}}
			\subfigure[]{\raisebox{-0.3cm}{\includegraphics[width=0.652\textwidth]{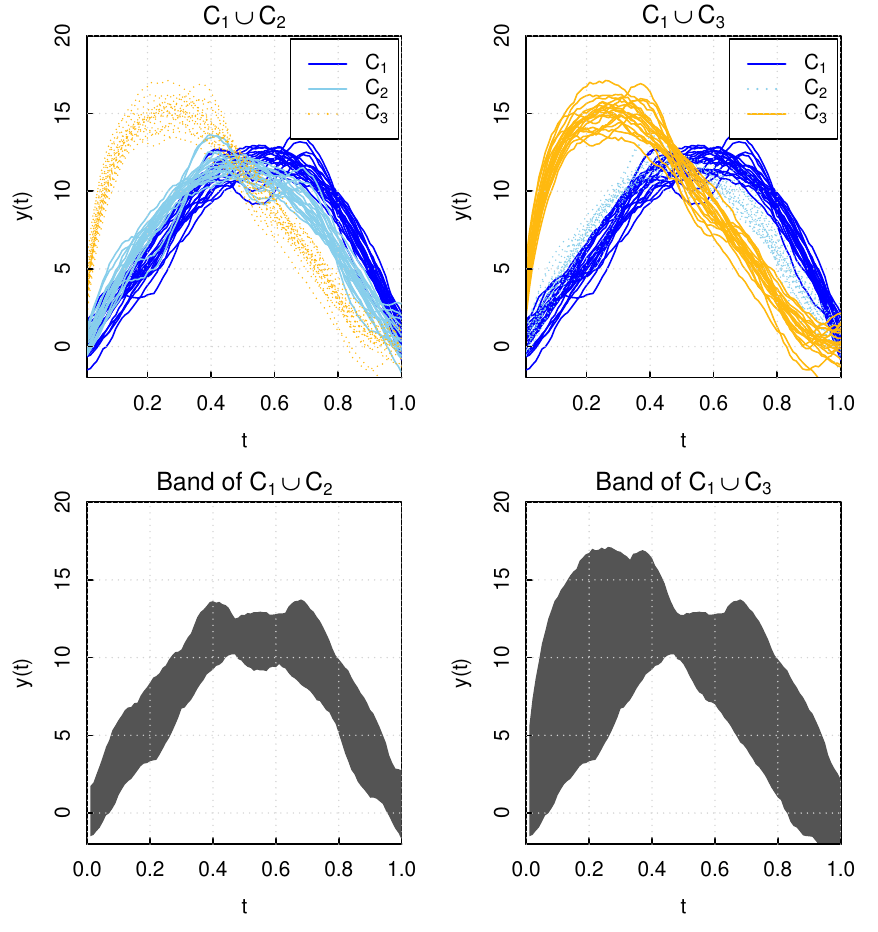}}}
		\end{center}
		\caption{(a) Ordinary Ward\textquotesingle s linkage and (b) the functional Ward-type linkage.}
		\label{ward}
	\end{figure}

	{ Analogous to the concept of within-cluster dispersion in ordinary
		Ward\textquotesingle s linkage in (\ref{eq1}), we define a functional
		Ward-type linkage criterion using the increase in band width to measure
		the dissimilarity between two functional-data clusters
		$\boldsymbol{C}_1$ and $\boldsymbol{C}_2$:
		\begin{equation}\label{eq2}
			D^2(\boldsymbol{C}_1,\boldsymbol{C}_2)
			=
			|\boldsymbol{C}_1\cup\boldsymbol{C}_2|
			W\!\left\{B(\boldsymbol{C}_1\cup\boldsymbol{C}_2)\right\}
			-
			|\boldsymbol{C}_1|
			W\!\left\{B(\boldsymbol{C}_1)\right\}
			-
			|\boldsymbol{C}_2|
			W\!\left\{B(\boldsymbol{C}_2)\right\}.
		\end{equation}
		Here, $B(\cdot)$ denotes the functional band generated by a cluster
		of curves and $W(\cdot)$ denotes its average width.
		
		Let $\mathcal I=[a,b]$ and let
		$\boldsymbol{C}=\{y_1,\ldots,y_n\}$. The functional band generated by
		$\boldsymbol{C}$ is defined as
		\[
		B(\boldsymbol{C})
		=
		\left\{
		(t,x)\in\mathcal I\times\mathbb R:
		\min_{1\leq r\leq n}y_r(t)
		\leq x
		\leq
		\max_{1\leq r\leq n}y_r(t)
		\right\}.
		\]
		Its average width is
		\[
		W\{B(\boldsymbol{C})\}
		=
		\frac{1}{b-a}
		\int_a^b
		\left\{
		\max_{1\leq r\leq n}y_r(t)
		-
		\min_{1\leq r\leq n}y_r(t)
		\right\}
		\,dt.
		\]
		Thus, $W\{B(\boldsymbol{C})\}$ is the average vertical distance between
		the upper and lower envelopes of the curves over the observation
		interval. A cluster containing highly homogeneous curves therefore
		has a relatively narrow functional band.}
	Figure \ref{ward} illustrates ordinary Ward\textquotesingle s linkage and the functional Ward-type linkage, showing that the linkage value between $\boldsymbol{C}_1$ and $\boldsymbol{C}_2$ is smaller than that between $\boldsymbol{C}_1$ and $\boldsymbol{C}_3$.
	
	To improve the robustness of clustering in the presence of atypical curves, we propose two functional Ward-type linkages built on the same principle. The linkage between two clusters is computed from a band-based dispersion measure, but the band is constructed using only a subset of the most central curves. Specifically, for a cluster $\bm{C}$, let $\mathcal{G}_{\tau}(\bm{C})\subseteq \bm{C}$ denote its $\tau$-central subset with $\tau\in(0,1]$, which is designed to exclude a small fraction of outlying observations and represent the typical behavior of the cluster. The two proposed linkages, termed magnitude-shape linkage (MS-linkage) and band-depth linkage (BD-linkage), differ only in how the central subset $\mathcal{G}_{\tau}(\bm{C})$ is identified. MS-linkage selects central curves based on a multivariate outlyingness measure, while BD-linkage selects them via a functional depth ranking.
	
	\subsection{Two Robust Linkages}\label{MSBD}
	MS-linkage and BD-linkage are built on the MS \citep{dai2019directional} and the MBD \citep{lopez2009concept,sun2011functional}, respectively. We first introduce the MS and MBD, and then describe how they are used to identify a subset of the most central curves.

	\subsubsection{MS and MBD}

	To construct the central subset for the MS-based linkage, we use
	directional outlyingness, which measures both the magnitude and the
	direction of a curve's deviation from the center.
	
	{  Let $F_{y(t)}$ denote the distribution of $y(t)$ at
		$t\in\mathcal I$. A statistical depth function measures the centrality
		of an observation relative to a distribution, with larger values
		indicating greater centrality. In this paper, we use projection depth
		for the pointwise depth underlying the directional outlyingness and
		MS-plot, following Dai and Genton
		\citeyearpar{dai2018multivariate,dai2019directional}. The calculations are
		performed using the \texttt{msplot} function in the
		\texttt{fdaoutlier} package.
		
		Let $z(t)$ denote the pointwise deepest value and define
		$
		d_y(t)=d\{y(t);F_{y(t)}\}.
		$
		
		The spatial sign from $z(t)$ to $y(t)$ is
		\[
		v_y(t)=
		\begin{cases}
			\dfrac{y(t)-z(t)}{|y(t)-z(t)|}, & y(t)\neq z(t),\\[1ex]
			0, & y(t)=z(t),
		\end{cases}
		\]
		and the pointwise directional outlyingness is
		$
		{\rm O}_y(t)
		=
		\left\{
		\frac{1}{d_y(t)}-1
		\right\}v_y(t).
		$
		In the univariate setting, $v_y(t)\in\{-1,0,1\}$.
		
		Dai and Genton \citeyearpar{dai2019directional} summarized directional
		outlyingness using
		$
		{\rm MO}(y)
		=
		\int_{\mathcal I}{\rm O}_y(t)w(t)\,dt
		$
		and
		$
		{\rm VO}(y)
		=
		\int_{\mathcal I}
		\left|
		{\rm O}_y(t)-{\rm MO}(y)
		\right|^2
		w(t)\,dt,
		$
		where $w(t)\geq0$ and
		$\int_{\mathcal I}w(t)\,dt=1$.
		The quantities $\mathrm{MO}(y)$ and $\mathrm{VO}(y)$ describe
		magnitude and shape outlyingness, respectively, and form the
		coordinates of the MS-plot. 
		
		{ The MS representation is not invariant to arbitrary separate
			rescaling of $\mathrm{MO}$ and $\mathrm{VO}$, because such rescaling
			changes their relative contribution and may alter the estimated
			two-dimensional quantile region. We therefore use the original
			MS coordinates without additional coordinatewise standardization.
			In contrast, projection depth is location--scale invariant: for
			$y^*(t)=a+b\,y(t)$ with $b>0$, directional outlyingness and hence
			$\mathrm{MO}$ and $\mathrm{VO}$ remain unchanged. Thus, the MS
			representation is invariant to positive location--scale transformations
			of the original curves, but not to arbitrary separate rescaling of its
			two coordinates.}
		
		Dai and Genton \citeyearpar{dai2018multivariate} proposed the MS-plot, a
		scatter plot of $(\mathrm{MO},\mathrm{VO})^\top$ that summarizes
		magnitude and shape outlyingness. Curves close to the center tend to
		have small $|\mathrm{MO}|$ and $\mathrm{VO}$; large
		$|\mathrm{MO}|$ mainly indicates a magnitude shift, whereas large
		$\mathrm{VO}$ indicates shape deviation. Curves nearer the center are
		therefore regarded as more representative, and a proportion $\tau$ is
		retained to construct the central band.}

	{ MBD provides a center-outward ordering of functional data. For
		$n$ curves $\{y_1,\ldots,y_n\}$ defined on a compact interval
		$\mathcal I=[a,b]$, let
		\[
		\mathrm{MBD}_n^{(j)}(y_i)
		=
		\binom{n}{j}^{-1}
		\sum_{1\le i_1<\cdots<i_j\le n}
		\frac{1}{b-a}
		\int_a^b
		\mathbf{1}
		\left\{
		\min_{r\in\{i_1,\ldots,i_j\}}y_r(t)
		\leq y_i(t)
		\leq
		\max_{r\in\{i_1,\ldots,i_j\}}y_r(t)
		\right\}
		\,dt.
		\]
		Thus, $\mathrm{MBD}_n^{(j)}(y_i)$ measures the average proportion of
		the observation interval over which $y_i$ lies within the band
		generated by the curves
		$y_{i_1},\ldots,y_{i_j}$. The modified band depth is then defined as
		\[
		\mathrm{MBD}_{n,J}(y_i)
		=
		\sum_{j=2}^{J}
		\mathrm{MBD}_n^{(j)}(y_i),
		\]
		where $J$ is a fixed integer satisfying $2\leq J\leq n$.}
	
	MBD is larger for curves that lie within many sample bands and
	therefore provides a natural center-outward ordering. { We set
		$J=2$ throughout for computational efficiency. Pairwise bands already
		provide a useful ordering, while the number of curve combinations
		increases rapidly for larger $J$. Because changing $J$ may change the
		MBD ranking and the selected central subset, $J=2$ should be viewed as
		a practical compromise rather than a claim of identical performance
		for all $J$.}
	
	Sun and Genton \citeyearpar{sun2011functional} used MBD to construct the
	functional boxplot, which summarizes the median, central envelope, and
	non-outlying range. Related extensions include surface, multivariate,
	trajectory, and sparse functional boxplots
	\citep{genton2014surface,dai2018functional2,yao2020trajectory,qu2022sparse}.
	In summary, the MS-plot decomposes outlyingness into magnitude and shape
	components, whereas MBD provides a scalar band-based centrality ranking.
	
	\subsubsection{MS-Linkage and BD-Linkage}
	
	For a cluster of $n$ curves
	$\bm{C}=\{y_1(t),y_2(t),\ldots,y_n(t)\}$,
	the subsets of the most central curves are denoted by
	$\mathcal{G}^{\rm MS}_{\tau}(\bm{C})$ and
	$\mathcal{G}^{\rm BD}_{\tau}(\bm{C})$, respectively.
	Specifically,
	\begin{itemize}
		\item MS-linkage selects curves corresponding to points in the
		lower central region of the MS-plot, which represents the most
		central points in the symmetrized point cloud
		$
		(\mathrm{MO},\mathrm{VO})^{\top}
		\cup
		(\mathrm{MO},-\mathrm{VO})^{\top}.
		$
		We apply the method proposed in Agarwal et al.
		\citeyearpar{agarwal2022flexible}, which is capable of handling both
		non-Gaussian and non-convex data, to calculate the
		two-dimensional (2D) $\tau$-quantile envelope of
		$
		(\mathrm{MO},\mathrm{VO})^{\top}
		\cup
		(\mathrm{MO},-\mathrm{VO})^{\top},
		$
		and identify the points lying within this envelope. The curves
		corresponding to these points form
		$\mathcal{G}^{\rm MS}_{\tau}(\bm{C})$.
		
		\item BD-linkage focuses on curves with large MBD values.
		We compute the MBD of each curve, yielding
		$
		\{\mathrm{MBD}(y_1),\ldots,\mathrm{MBD}(y_n)\}.
		$
		The $\lceil\tau n\rceil$ curves with the largest MBD values are
		retained to form $\mathcal G_{\tau}^{\mathrm{BD}}(\bm C).$
	\end{itemize}
	{
		For $r\in\{\mathrm{MS},\mathrm{BD}\}$, we first select the
		$\tau$-central subsets separately within the two candidate clusters.
		We then define their pooled central band as
		\[
		\mathcal{B}^{r}_{\tau}(\bm{C}_1,\bm{C}_2)
		=
		B\left(
		\mathcal{G}^{r}_{\tau}(\bm{C}_1)
		\cup
		\mathcal{G}^{r}_{\tau}(\bm{C}_2)
		\right).
		\]
		Thus, the pooled central band contains the central bands of both
		candidate clusters. This differs from recomputing a new central
		subset from $\bm{C}_1\cup\bm{C}_2$, which may discard the central
		representatives of one of the two original clusters.
		
		Accordingly, the MS-linkage is defined as
		\begin{align}
			D_{\rm MS}^2(\bm{C}_1,\bm{C}_2)
			={}&
			|\bm{C}_1\cup\bm{C}_2|
			\cdot
			W\left\{
			B\left(
			\mathcal{G}^{\rm MS}_{\tau}(\bm{C}_1)
			\cup
			\mathcal{G}^{\rm MS}_{\tau}(\bm{C}_2)
			\right)
			\right\}
			\nonumber\\
			&-
			|\bm{C}_1|
			\cdot
			W\left\{
			B\left(
			\mathcal{G}^{\rm MS}_{\tau}(\bm{C}_1)
			\right)
			\right\}
			\nonumber\\
			&-
			|\bm{C}_2|
			\cdot
			W\left\{
			B\left(
			\mathcal{G}^{\rm MS}_{\tau}(\bm{C}_2)
			\right)
			\right\},
			\label{eq3}
		\end{align}
		and the BD-linkage is defined as
		\begin{align}
			D_{\rm BD}^2(\bm{C}_1,\bm{C}_2)
			={}&
			|\bm{C}_1\cup\bm{C}_2|
			\cdot
			W\left\{
			B\left(
			\mathcal{G}^{\rm BD}_{\tau}(\bm{C}_1)
			\cup
			\mathcal{G}^{\rm BD}_{\tau}(\bm{C}_2)
			\right)
			\right\}
			\nonumber\\
			&-
			|\bm{C}_1|
			\cdot
			W\left\{
			B\left(
			\mathcal{G}^{\rm BD}_{\tau}(\bm{C}_1)
			\right)
			\right\}
			\nonumber\\
			&-
			|\bm{C}_2|
			\cdot
			W\left\{
			B\left(
			\mathcal{G}^{\rm BD}_{\tau}(\bm{C}_2)
			\right)
			\right\}.
			\label{eq4}
		\end{align}
		
		These quantities measure the weighted increase in band width required
		to construct a common band containing the central representatives of
		both clusters. They are nonnegative, as established below.
		
		\medskip
		\noindent
		\textbf{Proposition 1.}
		For $r\in\{\mathrm{MS},\mathrm{BD}\}$, the linkage criterion
		$D_r^2(\bm{C}_1,\bm{C}_2)$ defined in
		(\ref{eq3}) or (\ref{eq4}) satisfies
		$
		D_r^2(\bm{C}_1,\bm{C}_2)\geq 0
		$
		for any two nonempty and disjoint \footnote{Here, disjointness refers to observation indices rather than functional values, so distinct observations in the two clusters may still correspond to identical curves.
		} clusters $\bm{C}_1$ and
		$\bm{C}_2$ whose $\tau$-central subsets are nonempty.
		
		\medskip
		\noindent
		\textit{Proof.}
		For $r\in\{\mathrm{MS},\mathrm{BD}\}$, define
		\[
		W_{i,r}
		=
		W\left\{
		B\left(
		\mathcal{G}^{r}_{\tau}(\bm{C}_i)
		\right)
		\right\},
		\qquad i=1,2,
		\]
		and
		\[
		W_{12,r}
		=
		W\left\{
		B\left(
		\mathcal{G}^{r}_{\tau}(\bm{C}_1)
		\cup
		\mathcal{G}^{r}_{\tau}(\bm{C}_2)
		\right)
		\right\}.
		\]
		Since
		\[
		\mathcal{G}^{r}_{\tau}(\bm{C}_i)
		\subseteq
		\mathcal{G}^{r}_{\tau}(\bm{C}_1)
		\cup
		\mathcal{G}^{r}_{\tau}(\bm{C}_2),
		\qquad i=1,2,
		\]
		the band generated by the pooled central subsets contains the band
		generated by either individual central subset. Hence,
		\[
		W_{12,r}\geq W_{1,r}
		\qquad\text{and}\qquad
		W_{12,r}\geq W_{2,r}.
		\]
		Because the two clusters are disjoint,
		$|\bm{C}_1\cup\bm{C}_2|
		=|\bm{C}_1|+|\bm{C}_2|$. Therefore,
		\begin{align*}
			D_r^2(\bm{C}_1,\bm{C}_2)
			={}
			|\bm{C}_1|
			\left(W_{12,r}-W_{1,r}\right)
			+
			|\bm{C}_2|
			\left(W_{12,r}-W_{2,r}\right)
			\geq{}0.
		\end{align*}
		\hfill $\square$

		Although the notation $D_r^2$ is retained to emphasize its
		connection with Ward's linkage, the proposed quantity is used as a
		nonnegative cluster linkage criterion rather than as a metric.
		In particular, distinct clusters may generate the same central band,
		and neither the identity of indiscernibles nor the triangle inequality
		is required for hierarchical agglomerative clustering. Moreover, because the pooled band is constructed from central subsets
		selected separately within the two candidate clusters, the proposed
		criterion is not interpreted as the exact increment of a single
		within-cluster objective evaluated on the merged cluster. Rather, it
		measures the weighted additional band width required to construct a
		common band containing the robust representatives of the two clusters.}

\begin{figure}[htbp]
		\centering	
		\includegraphics[width=\textwidth]{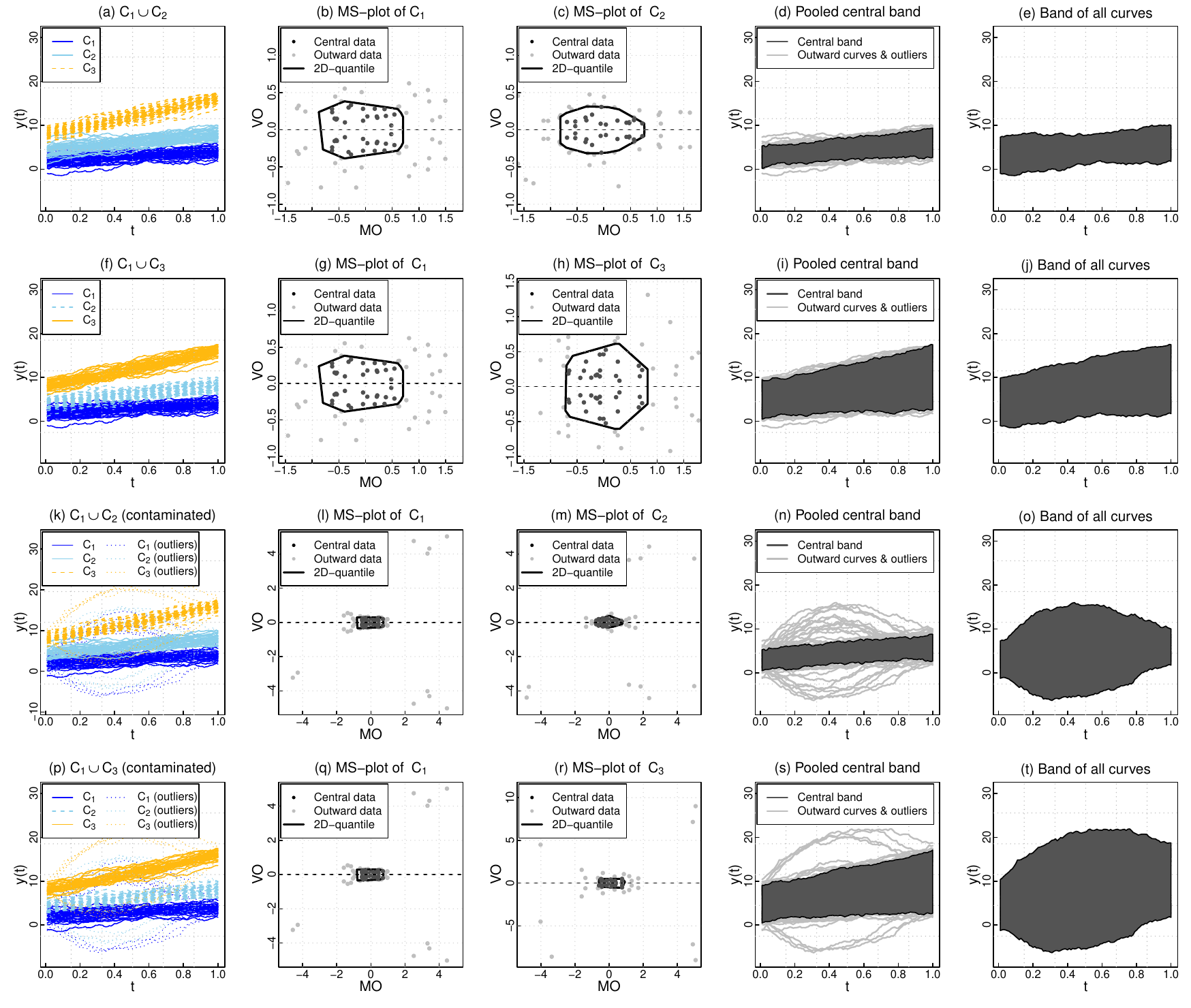}
		\caption{Illustration of the MS-linkage. The first two rows show the
			uncontaminated case and the last two rows the contaminated case. Rows 1
			and 3 compare $\boldsymbol{C}_1$ with $\boldsymbol{C}_2$, while Rows 2
			and 4 compare $\boldsymbol{C}_1$ with $\boldsymbol{C}_3$. The columns
			show the curves, the separately constructed symmetrized MS-plots for the two
			candidate clusters, the pooled central band, and the band generated by
			all curves, respectively. The central-subset proportion is $\tau=0.5$.}
		\label{fig-movo1}
	\end{figure}

\begin{figure}[htbp]
		\centering
		\includegraphics[width=0.63\textwidth]{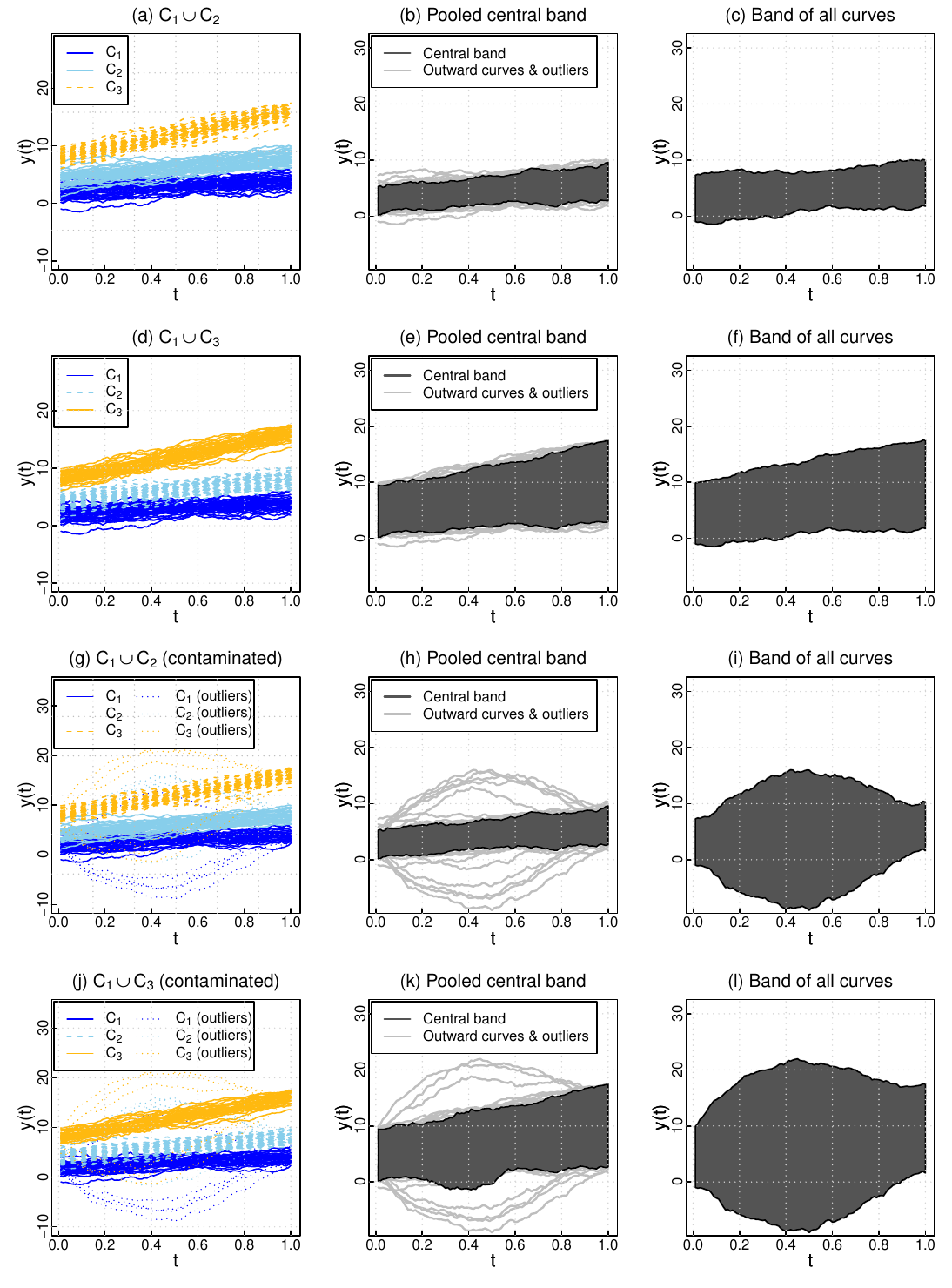}
		\caption{
			Illustration of the BD-linkage. The first two rows show the uncontaminated
			case, while the last two rows show the corresponding contaminated case.
			Rows 1 and 3 compare $\boldsymbol{C}_1$ with $\boldsymbol{C}_2$, whereas
			Rows 2 and 4 compare $\boldsymbol{C}_1$ with $\boldsymbol{C}_3$.
			The first column displays the three clusters, with the candidate pair
			indicated in each panel title.
			The second column shows the pooled central band generated by the union
			of the BD-central subsets selected separately within the two candidate
			clusters, with the remaining outward curves and outliers shown in light
			gray. The third column shows the band generated by all curves.
			The central-subset proportion is $\tau=0.5$.
		}
		\label{fig-fb}
	\end{figure}

	Figure~\ref{fig-movo1} illustrates the MS-linkage. In the
	uncontaminated case, $\boldsymbol{C}_1$ and $\boldsymbol{C}_2$ have a
	narrower pooled central band than $\boldsymbol{C}_1$ and
	$\boldsymbol{C}_3$. Under 20\% contamination (Model~2 in
	Section~\ref{SM}), the pooled MS-central bands change only moderately,
	whereas the bands generated by all curves expand substantially.
	Figure~\ref{fig-fb} shows the same comparison for the BD-linkage:
	the pooled BD-central bands remain relatively stable after
	contamination, while the all-curve bands change markedly. These
	illustrations show how central-subset selection limits the influence of
	outlying curves on both linkage criteria.

	\subsection{Clustering Algorithm}
\begin{figure}[htbp]
		\centering
		\includegraphics[width=0.95\textwidth]{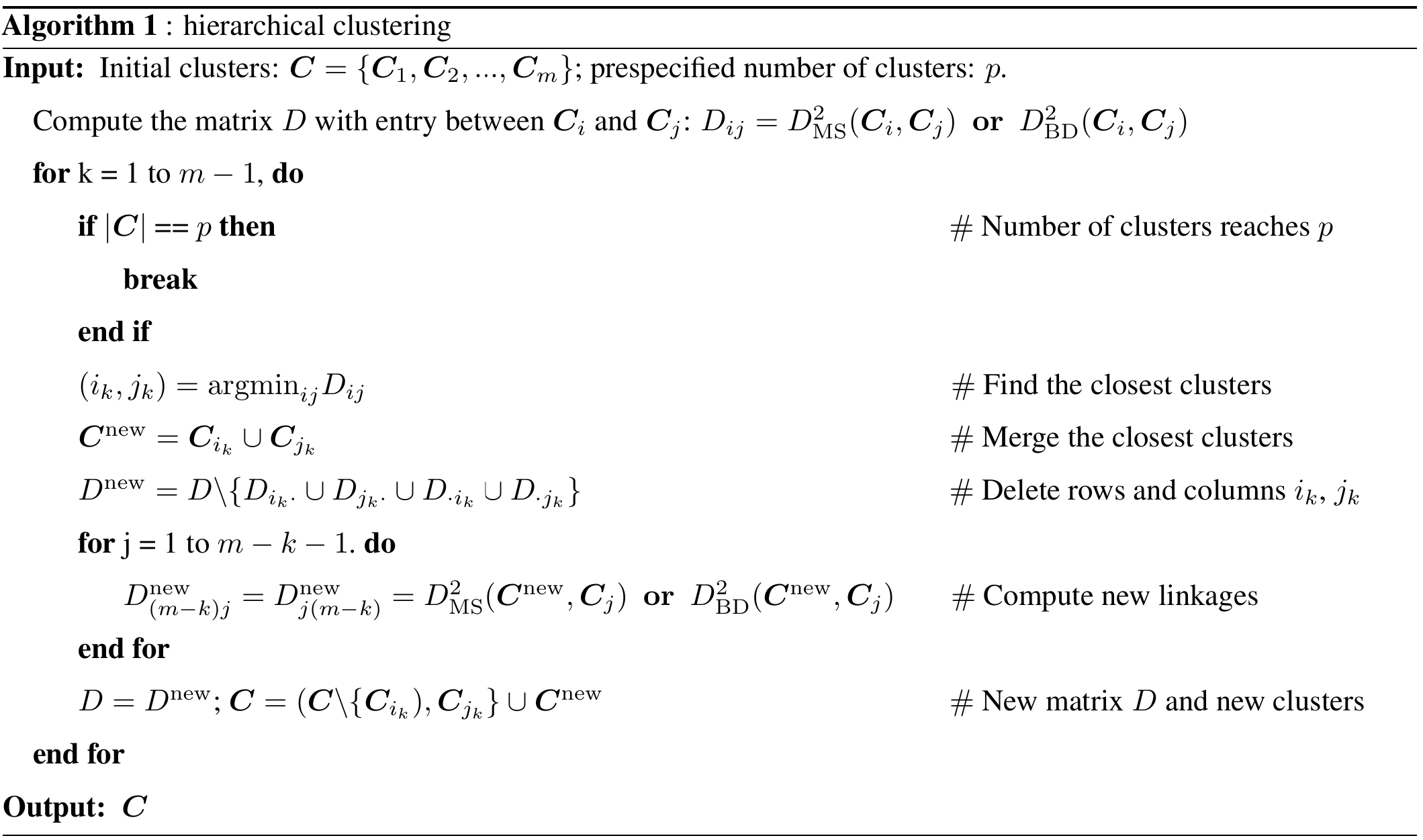}
		\caption{Algorithm of clustering.}
		\label{algy}
	\end{figure}

	Suppose we have $m$ initial clusters, the hierarchical clustering algorithm proceeds as follows:
	\noindent \\
	{\bf Step 1}: Let $\{\boldsymbol{C}_1,\boldsymbol{C}_2,\ldots,\boldsymbol{C}_m\}$ be the initial clusters.\\
	{\bf Step 2}: Compute the initial linkage matrix based on criterion (\ref{eq3}) or (\ref{eq4}).\\
	{\bf Step 3}: Identify the two clusters with the smallest linkage value,
	merge them, and update the linkage matrix by computing only the
	linkages involving the newly formed cluster. \\
	{\bf Step 4}: Repeat Step~3 until the number of clusters is reduced to one
	or reaches a prespecified number of clusters.
	Algorithm 1 in Figure \ref{algy} provides the pseudocode and summarizes the procedure for hierarchical clustering.
	
	\noindent
	\Needspace{5\baselineskip}
	{\bf Remarks.} 
	\begin{enumerate}
		\item Sample size issue: If an initial cluster contains too few curves
		to reliably identify the most central curves, additional functional
		replicates may be obtained when permitted by the natural observational
		structure. For example, in Section~4, long time series are divided into
		shorter segments so that multiple spectral curves from the same
		recording form one initial cluster. Further details on small-cluster
		handling are provided in
		Section \ref{initial_size}.
		\item Determining the number of clusters: The predefined number of clusters, $p$, can be chosen based on domain expertise or determined using data-driven methods such as the elbow method, the Calinski-Harabasz index \citep{calinski1974dendrite}, or the approaches described in Chapter 17 of Gan et al. \citeyearpar{gan2020data}.
	\end{enumerate}
	\subsection{Initial-Cluster Size and Tuning Parameter Selection}
	\subsubsection{Initial-Cluster Size}
	\label{initial_size}
	
	{ The construction of the central subset requires a sufficient number of
		curves within each cluster. When only a few curves are available, the
		centrality ordering and the resulting central band may become unstable.
		Based on our empirical experience, the standard BD-based central-subset
		construction is reasonably stable when a cluster contains at least four
		curves, whereas the standard MS-based construction generally requires
		at least twelve curves. These values should be interpreted as practical
		recommendations for stable central-subset estimation rather than as
		strict theoretical lower bounds.
		
		Whenever the observational structure permits, we therefore recommend
		forming initial clusters that are sufficiently large. In applications
		involving long time series, one practical strategy is to divide each
		series into several shorter segments and compute one functional summary
		from each segment. The curves obtained from the same recording can then
		be treated as one initial cluster. This strategy is used in
		Section~\ref{APP}, where segmentation increases the number of spectral
		curves available within each recording-level cluster.
		
		To ensure that the proposed linkages remain well defined when a current
		cluster contains fewer curves than the recommended sizes, we adopt the
		following small-cluster rule. Let $\bm C$ denote a current cluster and
		let $\widetilde{\bm C}^{\,r}$ denote its representative subset for
		$r\in\{\mathrm{MS},\mathrm{BD}\}$.
		\begin{enumerate}
			\item If $|\bm C|=1$, the single curve is retained:
			$
			\widetilde{\bm C}^{\,r}=\bm C.
			$
			
			\item If $|\bm C|=2$, both curves are retained:
			$
			\widetilde{\bm C}^{\,r}=\bm C.
			$
			
			\item If $|\bm C|=3$, the BD-based procedure retains the two curves
			having the largest modified band-depth values, whereas the MS-based
			procedure retains the two curves closest to the origin in the
			magnitude--shape plot.
			
			\item If $4\leq |\bm C|\leq 11$, the BD-based procedure applies its
			usual depth-based central-subset selection. For the MS-based
			procedure, the representative subset is determined using the
			quadrilateral central region around the origin of the
			magnitude--shape plot. If the quadrilateral contains fewer than two curves, we retain the two
			curves closest to the origin of the MS-plot.
			
			\item If $|\bm C|\geq 12$, the usual central-subset construction
			based on the specified value of $\tau$ is used for both the MS- and
			BD-based procedures.
		\end{enumerate}
		
		After the representative subsets have been determined separately for
		two current clusters $\bm C_1$ and $\bm C_2$, their linkage is evaluated
		from the pooled band generated by
		$
		\widetilde{\bm C}_1^{\,r}
		\cup
		\widetilde{\bm C}_2^{\,r}.
		$
		Thus, the same linkage definition is used at every stage of the
		agglomerative procedure, while the small-cluster rule provides a
		well-defined representative subset when the ordinary centrality-based
		selection is not sufficiently stable.
		
		In general, larger initial clusters are preferable because they contain
		more functional information for identifying representative central
		curves. When an initial cluster contains only a few curves, the
		centrality ranking and the resulting central-band estimate become more
		variable, and in extreme cases a reliable central subset may not be
		obtained. This behavior is also supported by the sensitivity analysis in the
		simulation study, where clustering performance becomes less stable as
		the initial-cluster size is reduced, particularly near the practical
		threshold of the MS-based construction. Moreover, for a fixed total number of curves, increasing the
		initial-cluster size reduces the number of initial clusters and hence
		the number of pairwise linkage evaluations, leading to a lower overall
		computational cost. In practice, however, sufficiently large initial
		clusters may not always be available. For example, a time series may be
		too short to generate many segments that can be grouped into the same
		initial cluster. Therefore, whenever the data structure permits, we
		recommend using relatively large initial clusters, while respecting the
		natural observational grouping of the data.}
	
	\subsubsection{Selection of the Central Proportion \texorpdfstring{$\tau$}{tau}}
	\label{tau}
	{ The tuning parameter $\tau\in(0,1]$ determines the proportion of
		curves retained in the central subset of each cluster. A smaller value
		of $\tau$ places greater emphasis on the most central curves and
		provides stronger protection against within-cluster contamination,
		whereas a larger value retains more information about the natural
		functional variation within the cluster.
		
		In the numerical studies of this paper, $\tau$ is specified in advance
		and is not tuned separately for individual simulation settings,
		contamination mechanisms, Monte Carlo replications, or real datasets.
		In all main simulation experiments, we fix $\tau=0.5$ in advance,
		except for the sensitivity analysis in Section \ref{sen}, where $\tau$
		is deliberately varied to assess its influence.
		Thus, the linkage is constructed using the most central half of the
		curves in each cluster. This choice is consistent with the conventional
		central-region interpretation used in functional boxplots and gives
		substantial protection against the contamination mechanisms considered
		in the simulations. The sensitivity analysis in
		Figure~\ref{ratio} further shows that $\tau=0.5$ provides stable
		clustering performance over the settings examined.
		
		For all real-data applications, we fix
		$
		\tau=0.7.
		$
		The real datasets do not provide known contamination proportions or
		labels identifying contaminated curves. We therefore use a moderately
		larger central proportion to retain more of the genuine within-group
		functional variability and reduce the possibility of excluding valid
		but relatively peripheral curves. At the same time, $\tau=0.7$
		continues to exclude the most peripheral 30\% of curves from the
		central-band construction. The same value is used for all real-data
		applications and is not selected by optimizing the reported clustering
		results.
		
		For applications in which a data-dependent value of $\tau$ is desired,
		two optional strategies may be considered. The first is based directly
		on clustering stability or internal validity. Specifically, one may
		evaluate a grid $\mathcal T$ of candidate values and examine the
		agreement of the resulting partitions under subsampling, bootstrap
		resampling, or small data perturbations. A value of $\tau$ associated
		with high and stable agreement may then be selected. Internal
		clustering validity measures, such as the silhouette coefficient or
		Dunn index, may also be used as supplementary diagnostics.
		
		The second is a trimming-inspired strategy. The retained proportion
		$\tau$ is heuristically related to the retained fraction in
		trimming-based clustering. Consequently, procedures developed for
		selecting a trimming proportion, such as classification trimmed
		likelihood curves and parametric bootstrap diagnostics
		\citep{fritz2012tclust,garcia2025choice,torti2021semiautomatic,
			rivera2019robust}, may provide additional guidance. This relationship
		is only heuristic because trimming and within-cluster central-subset
		selection are different robustness mechanisms. Neither of these
		data-dependent strategies is used in the numerical analyses of the
		present paper.}

	\subsection{Computational Complexity}
	\label{comp}
	{
		The proposed MS- and BD-linkages are computationally more demanding
		than ordinary Ward\textquotesingle s linkage. Ordinary Ward clustering
		can update the within-cluster sum of squares, or the corresponding
		cluster dissimilarities, recursively after each merge. In contrast,
		the proposed procedures require the representative central subset of
		each newly formed cluster to be determined before its linkages to the
		remaining clusters are evaluated. The additional computational cost
		therefore arises mainly from repeated centrality ranking, central-band
		construction, and linkage-matrix updating.
		
		Let $m$ denote the number of initial clusters, let $m_0$ denote the
		number of curves in each initial cluster, and let $T$ denote the number
		of sampling points per curve. The total number of curves is therefore
		$
		N=mm_0.
		$
		The agglomerative procedure performs at most $m-1$ merging steps. At
		each step, the central subset of the newly formed cluster is computed,
		and its linkage values with the remaining clusters are evaluated.
		
		For the BD-linkage, the cost of constructing a central subset can be
		described more explicitly. Consider a current cluster containing $n$
		curves. Because we use $J=2$, the pairwise MBD values can be evaluated
		from the pointwise ranks of the curves. At each sampling point, sorting
		the $n$ functional observations requires $O(n\log n)$ operations.
		Repeating this calculation over all $T$ sampling points gives
		$
		O(Tn\log n)
		$
		operations for computing the MBD ordering. Selecting the
		$\lceil\tau n\rceil$ most central curves and constructing their
		pointwise lower and upper envelopes require at most $O(n\log n)$ and
		$O(Tn)$ additional operations, respectively. These terms do not change
		the leading order, so the overall cost of constructing the BD-central
		subset and its band is approximately
		$
		O(Tn\log n).
		$
		
		To obtain a conservative worst-case bound, suppose that at a particular
		stage there are $q$ current clusters with sizes
		$n_1,\ldots,n_q$, where
		$
		\sum_{a=1}^{q}n_a=N.
		$
		Under a straightforward implementation in which the centrality
		information is recomputed when the candidate linkage values are
		evaluated, each current cluster may be involved in $q-1$ candidate
		pairs. The computational cost at this stage is therefore bounded by
		$
		O\left[
		T(q-1)
		\sum_{a=1}^{q}n_a\log n_a
		\right].
		$
		Since $n_a\leq N$ for every $a$,
		$
		\sum_{a=1}^{q}n_a\log n_a
		\leq
		N\log N,
		$
		and hence the cost at this stage is bounded by
		$
		O(TqN\log N).
		$
		Summing over the agglomerative stages, for which $q$ decreases from
		$m$ to $2$, gives
		\[
		O\left\{
		TN\log N
		\sum_{q=2}^{m}q
		\right\}
		=
		O(TNm^2\log N).
		\]
		Substituting $N=mm_0$ yields the conservative worst-case upper bound
		\[
		O\left\{
		T\,m^3m_0\log(mm_0)
		\right\}.
		\]
		
		This upper bound corresponds to extensive recalculation and is
		deliberately conservative. In the actual implementation, the central
		subsets and band envelopes of unchanged clusters can be stored and
		reused. After two clusters are merged, only the central subset of the
		newly formed cluster and the corresponding row and column of the
		linkage matrix need to be updated. If $h_s$ denotes the number of
		curves in the cluster produced at the $s$th merge, the update-based
		computational workload can be expressed more precisely as
		$
		O\left\{
		T\sum_{s=1}^{m-1}h_s\log h_s
		+
		Tm^2
		\right\},
		$
		where the first term represents centrality ranking for newly formed
		clusters and the second represents the pairwise linkage updates. In
		typical applications, most calculations occur before clusters have
		grown close to the total size $N$, and the current cluster sizes remain
		closer to the initial size $m_0$. A useful practical approximation is
		therefore
		\[
		O\left\{
		T\,m^2m_0\log m_0
		\right\}.
		\]
		This expression should be interpreted as a practical workload
		approximation rather than as a uniform asymptotic upper bound.
		
		The use of $J=2$ also plays an important role in controlling the
		computational cost of the BD-linkage. Pairwise bands permit a
		rank-based calculation of MBD, whereas the number of possible bands
		increases rapidly when higher-order combinations are included. In
		particular, direct evaluation with a general value of $J$ may involve
		up to
		$
		\sum_{j=2}^{J}\binom{n}{j}
		$
		curve combinations within a cluster. Thus, setting $J=2$ avoids the
		rapidly increasing cost associated with higher-order bands and
		provides a practical compromise between computational efficiency and
		the amount of band information used.
		
		The MS-linkage is generally more computationally demanding than the
		BD-linkage. In addition to calculating pointwise directional
		outlyingness and constructing the MS-plot, it requires estimation of a
		two-dimensional $\tau$-quantile contour from the symmetrized MS-plot
		point cloud. The non-convex contour procedure used in this paper
		\citep{agarwal2022flexible} introduces additional computational
		overhead. Its cost depends not only on the number of curves in the
		current cluster but also on the geometric complexity of the estimated
		contour, including the number of boundary arcs or polygonal segments.
		Consequently, a single general complexity expression for the
		MS-linkage is less informative than the preceding bound for the
		BD-linkage.
		
		Several strategies may be considered to reduce the computational cost
		for large functional datasets. First, the central subsets, central-band
		envelopes, and unchanged linkage values can be cached so that only the
		newly formed cluster is processed after each merge. Second, the
		two-dimensional quantile contour used by the MS-linkage may be
		simplified by reducing the number of boundary segments or by
		approximating it with a simpler polygon, such as a rectangle. More
		generally, a penalty on the number of contour arcs may be introduced to
		balance geometric flexibility and computational efficiency. Such
		simplifications may, however, change the selected central subset and
		potentially reduce clustering accuracy. They are therefore possible
		extensions rather than procedures used in the numerical studies of
		this paper.}
	
	\section{Simulations}\label{SM}
	We use two experiments to evaluate the proposed linkages. The first experiment examines their robustness to outlier contamination models considered in related work. The second experiment is designed to test whether the linkages can effectively handle several commonly encountered artifacts in EEG data, such as eye-blinks and eye-movements.
	\subsection{Experiment Design}
	For the two experiments, we consider $20$ initial clusters, each containing $30$ curves. There are $4$ true clusters (ground truth)  with $k=1,\ldots,4$ denoting the $k$-th true cluster, where each true cluster consists of five initial clusters. Each curve has a probability of being contaminated with an outlier, and we choose three contamination rates $c = 0.1, 0.15, 0.2$. The data are sampled at $T = 100$ time points over the interval $\mathcal{I} = [0,1]$. { We use $\tau=0.5$ throughout Experiments~1 and~2; $\tau$ is varied
		only in the sensitivity analysis.}
	
	\subsubsection{Experiment 1: Outlier Models}
\begin{figure}[htbp]
		\centering
		\includegraphics[width=1\textwidth]{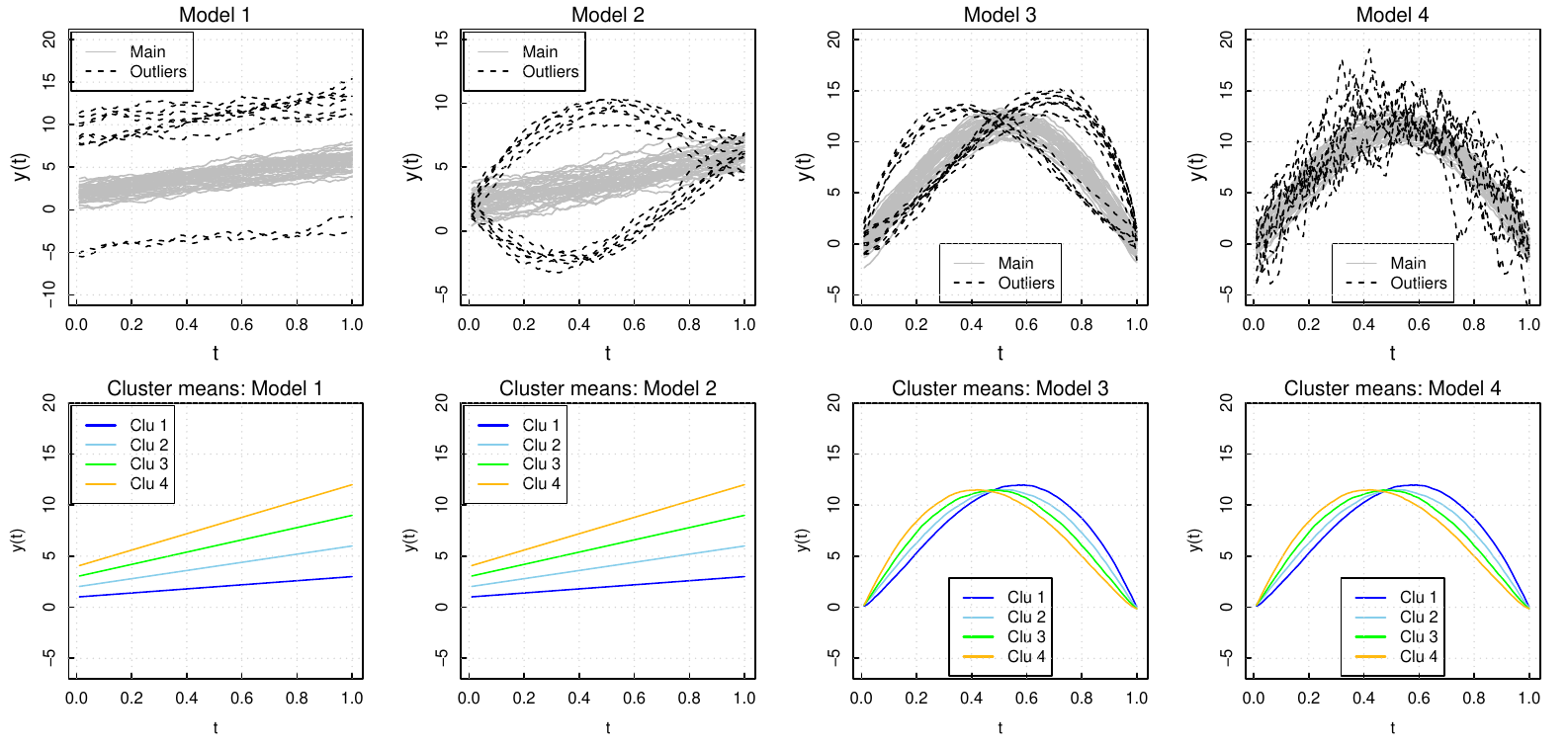}
		\caption{Top: illustrative examples of the four outlier models. The contamination rate is $c=0.2$; bottom:  the cluster means of the four true clusters.}
		\label{fig-out}
	\end{figure}

	In the first experiment, we consider the outlier models used in López-Pintado and Romo \citeyearpar{lopez2009concept}, Dai and Genton \citeyearpar{dai2018multivariate}, and Dai and Genton \citeyearpar{dai2019directional}.  Details of the models are described as follows:
	\vspace{-0.3cm}
	\begin{itemize}
		\item {\bf Model 1}\\
		Main model: $y(t) = k + 2kt + e_1(t)$,\\
		Contamination model: $y(t) = k + 2kt + 8U + e_1(t)$ with contamination probability $c$,\\	
		for $0\leq t \leq 1$, where $e_1(t)$ is a Gaussian process with zero mean and covariance function $\gamma(s,t) = {\rm exp}\{-|s-t|\}$, and $U$ takes values -1 and 1 with equal probability. The contaminating curves shift up and down from the main model.

		\item {\bf Model 2}\\
		Main model: $y(t) = k + 2kt + e_1(t)$,\\
		Contamination model: $y(t) = k + 2kt + Ug(t) + e_1(t)$ with contamination probability $c$,\\
		for $0\leq t \leq 1$, where $g(t) = 60t^{1.5}(1-t)$. The contaminating curves alter the shape of the main model.
		
		\item {\bf Model 3}\\
		Main model: $y(t) = 60(1-t)^{1+0.125(k-1)}t^{1.375-0.125(k-1)}+ e_2(t),$\\
		Contamination model: $y(t) = 60(1-t)^{1+0.125(k-1+2U)}t^{1.375-0.125(k-1+2U)}+ e_2(t)$ with contamination probability $c$,\\
		for $0\leq t \leq 1$ and $e_2(t)$ is a Gaussian process with zero mean and covariance function $\gamma(s,t) = {\rm exp}\{-0.5|s-t|\}$. The peaks of contaminating curves shift left and right from the main model.
		
		\item {\bf Model 4}\\
		Main model: $y(t) = 60(1-t)^{1+0.125(k-1)}t^{1.375-0.125(k-1)}+ e_{3,k}(t),$\\
		Contamination model: $y(t) = 60(1-t)^{1+0.125(k-1+2U)}t^{1.375-0.125(k-1+2U)}+ e_4(t)$ with contamination probability $c$,\\
		for $0\leq t \leq 1$, $e_{3,k}(t)$ is a Gaussian process with zero mean and covariance function $\gamma(s,t) = {\rm exp}\{-0.5k^2|s-t|\}$,  and $e_4(t)$ is a Gaussian process with zero mean and covariance function $\gamma(s,t) = 8 {\rm exp}\{-|s-t|/0.05\}$. The peaks of contaminating curves shift left and right from the main model. The contaminating curves share the same trend as the main model, but exhibit distinct covariance structures that lead to large oscillations.
	\end{itemize}
	Figure~\ref{fig-out} illustrates the four models: the top row shows the curves and outliers within a representative initial cluster, while the bottom row displays the cluster means (without contamination) to show the between-cluster mean differences.
	
	For model 4, we cluster the data in the spectral domain using smoothed log-periodograms to represent the spectral features of the time series. The covariance structures in each cluster are also different, which leads to different spectral features. Although raw periodograms are unbiased estimators of the spectral density functions (SDFs), they suffer from their roughness and high variance. As a result, smoothing is required to produce consistent estimators. In this paper, the smoothing bandwidth is selected automatically using the gamma generalized cross-validation (gamma-GCV) method \citep{ombao2001simple}. The log-periodograms introduce bias when estimating log-SDFs due to Jensen’s inequality. To correct for this, we apply a bias adjustment by adding the Euler-Mascheroni constant  $\gamma=0.57721$ to the smoothed log-periodograms \citep{wahba1980automatic}.  
	
	\subsubsection{Experiment 2: EEG Artifacts}
\begin{figure}[htbp]
		\begin{center}
			\subfigure[]{\raisebox{0.4cm}{\includegraphics[width=0.72\textwidth]{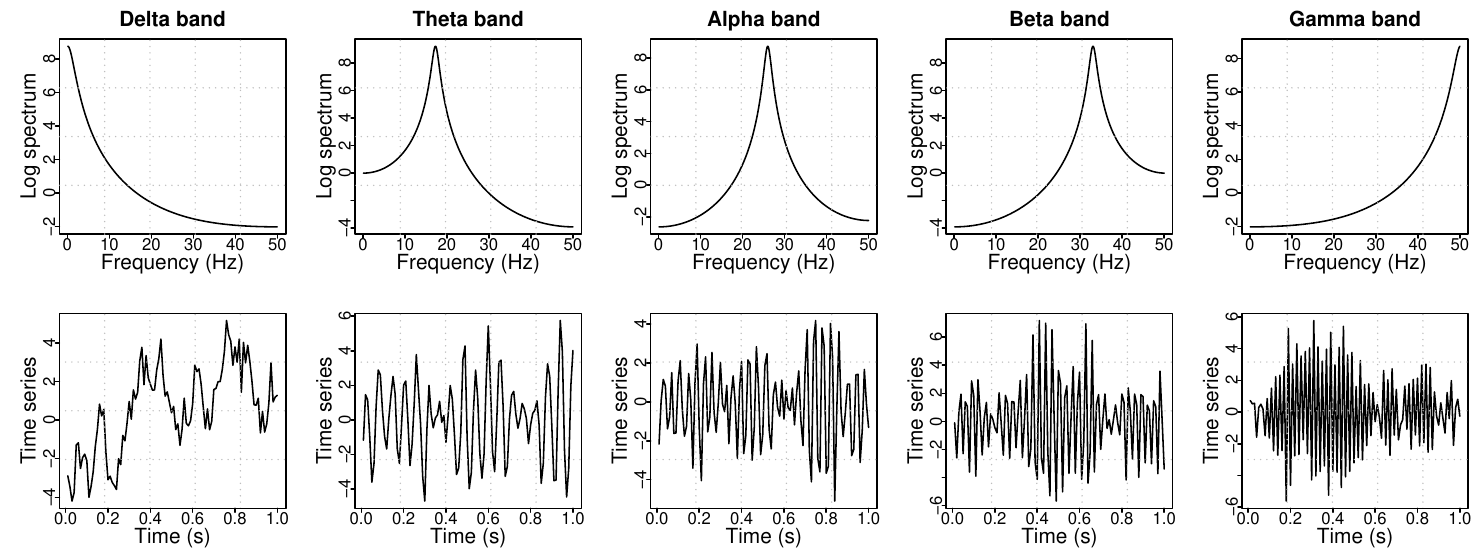}}}
			\subfigure[]{\includegraphics[width=0.8\textwidth]{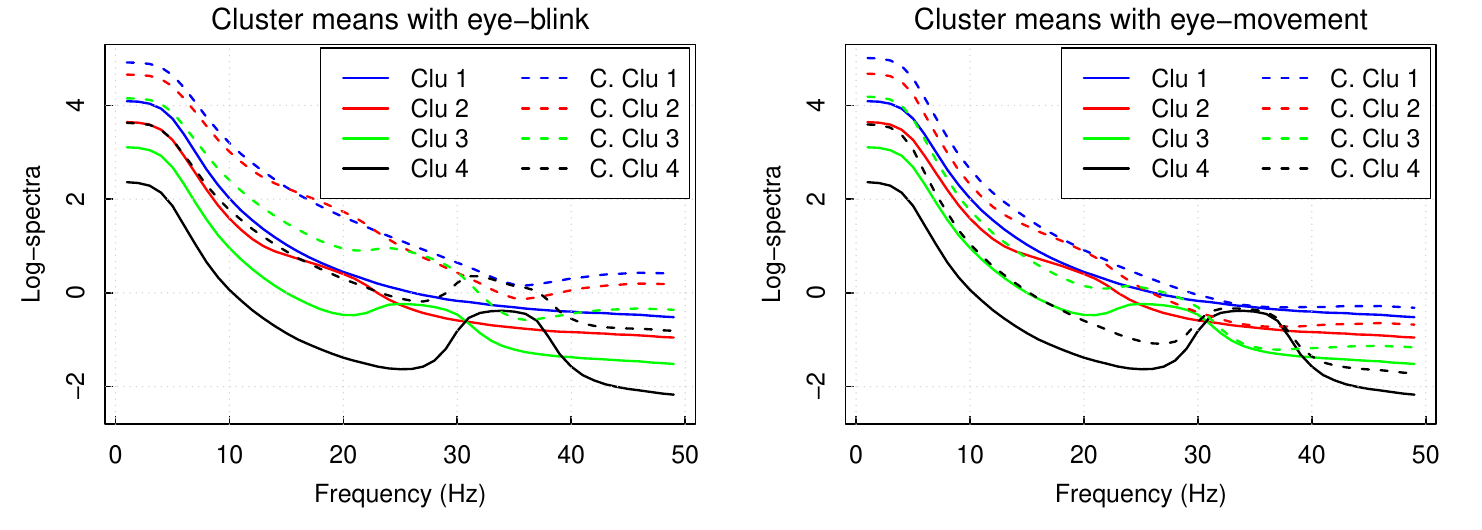}}
			\subfigure[]{\includegraphics[width=.8\textwidth]{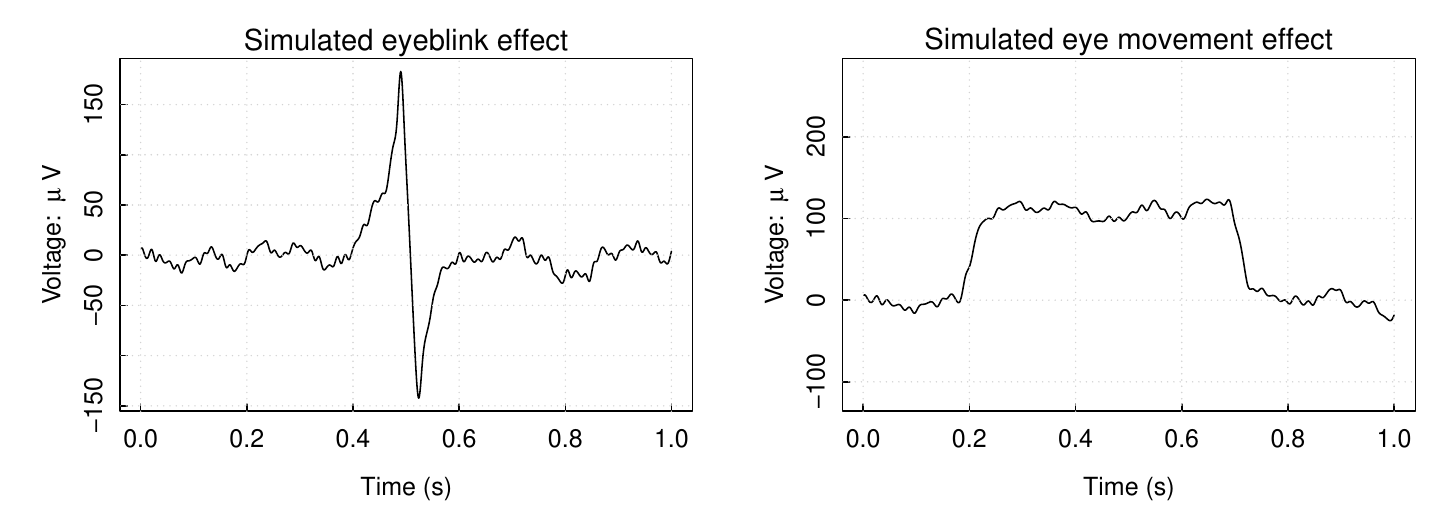}}
		\end{center}
		\caption{(a) The log-SDFs of the five latent sources and the corresponding realizations. (b) The averaged log-SDFs of the four clusters and the contaminated clusters ($c = 0.2$); left panel: eye-blink, and right panel: eye-movement. “C.” denotes contaminated. Solid lines represent the uncontaminated cluster means, whereas
			long-dashed lines represent the corresponding contaminated cluster means. (c) The simulated artifacts.}
		\label{spec}
	\end{figure}

	In the second experiment, we simulate EEG data with artifacts and cluster the data in the frequency domain. EEG is widely used in both laboratory and clinical applications due to its non-invasiveness, affordability, portability, and high temporal resolution. We use a mixture of second-order autoregressive $(AR(2))$ models to capture the oscillatory activities of EEG signals:
	$$y_t=\phi_1 y_{t-1}+\phi_2 y_{t-2}+w_t,$$ 
	where $w_t$ is the white-noise process with variance $\sigma_w^2$. For an AR(2) process with a pair of complex-conjugate roots, let
	the roots be denoted by $z_1$ and $z_2$, where $M>1$. The magnitude $|z_1|=|z_2|=1/r > 1$ ensures the causality. The AR coefficients $\phi_1$ and  $\phi_2$ determine the peak frequency $\omega_0$ in the SDF: 
	$$
	\phi_1 = \frac{2 \cos (2\pi \omega_0 / F)}{M} \text{ } \text{ }\text{ }\text{ } {\rm and} \text{ }\text{ }\text{ }\text{ } \phi_2=-\frac{1}{M^2} , 
	$$
	where $F$ is the sampling frequency in Hertz. $M$ controls the peak sharpness, with the peak becoming narrower as $M \rightarrow 1^+$. Then, by setting appropriate AR coefficients, one can generate time series with specific spectral features. Clinical EEG analysis usually focuses on five frequency bands, and we generate the corresponding sources using AR(2) models with different AR coefficients.
	Details about the five latent sources are shown in Table \ref{tab:eeg_ar2}. These coefficients are used directly in the simulation and are not
	required to satisfy the above complex-root parameterization in every
	case. To illustrate the oscillatory patterns within each frequency band, Figure \ref{spec}(a) depicts the log-SDFs and examples of the generated time series, where the oscillations become more rapid as the peak frequency approaches $50$ Hz.
	
	\begin{table}[ht]
		\renewcommand{\arraystretch}{1.3} 
		
		\centering
		\caption{Five latent AR(2) sources used for EEG signals simulation.}
		\begin{tabular}{ccc}
			\toprule[1.2pt]
			Frequency Band &  Frequency (Hz) & AR Coefficients $(\phi_1, \phi_2)$ \\
			\hline
			Delta   & 0 -- 4 Hz  & (0.8, -0.1)  \\
			Theta   & 4 -- 8 Hz  & (0.9, -0.9) \\
			Alpha  & 8 -- 16 Hz & (-0.1, -0.9) \\
			Beta  & 16 -- 32 Hz & (-0.9, -0.9) \\
			Gamma  & 32 -- 50 Hz & (-0.8, -0.1) \\
			\toprule[1.2pt]
		\end{tabular}
		
		\label{tab:eeg_ar2}
	\end{table}
	
	The EEG data, simulated by mixture AR(2) models, are generated as linear combinations of the five latent sources. The log-SDFs of the four true clusters are:
	$$
	f_1 = \log\left(\frac{4}{5} \tilde{f}_1  + \frac{1}{10}  \tilde{f}_2\right),\text{ }\text{ }
	f_2 = \log\left(\frac{3}{5}  \tilde{f}_1 + \frac{1}{10}  \tilde{f}_3\right),$$
	$$f_3 = \log\left(\frac{2}{5}  \tilde{f}_1 + \frac{1}{10}  \tilde{f}_4\right),\text{ }\text{ }
	f_4 = \log\left(\frac{1}{5}  \tilde{f}_1 + \frac{1}{10}  \tilde{f}_5\right),
	$$
	where $ \tilde{f}_i(\omega), i=1,\ldots,5$ are the SDFs of the five latent AR(2) sources.  The cluster mean functions (with and without contamination) are shown in Figure~\ref{spec}(b), illustrating how outliers shift the mean within each cluster.

	We consider two types of clinical artifacts as outliers: eye-blink and eye-movement effects.  EEG can be contaminated in the frequency or time domain by these artifacts, which result from internal sources such as physiological activities and subject movement, as well as external sources including environmental interference, equipment malfunction, and the movement of electrodes and cables. In our setting, eye-blinks are simulated by the difference of two gamma functions, while eye-movements are generated by adding a wide peak to the EEG signal. The shape of the simulated artifacts is based on clinical EEG data in Mansor et al. \citeyearpar{mansor2011integrating} and Abo-Zahhad et al. \citeyearpar{abo2015new}. Sazgar et al. \citeyearpar{sazgar2019eeg} and Mumtaz et al. \citeyearpar{mumtaz2021review} provide a comprehensive introduction of different types of EEG artifacts. The simulated artifacts are shown in Figure \ref{spec}(c).

	\subsection{Measures of Quality}
	To evaluate the performance of the proposed linkages in discovering the correct labels in a more systematic framework, we use two popular measures: the adjusted Rand index (ARI) \citep{vinh2009information} and the SIM index in the \texttt{R} package \texttt{TSclust} \citep{montero2015tsclust}, which are commonly used in the clustering evaluation literature. 
	
	{ The adjusted Rand index (ARI) measures the agreement between the
		estimated clustering partition and the true class labels while
		adjusting for agreement that may occur by chance. An ARI value of
		$1$ indicates perfect agreement, a value close to $0$ indicates
		agreement comparable to random assignment, and a negative value
		indicates agreement worse than that expected by chance. Larger ARI values therefore represent better clustering performance.}

	We also use the SIM index, which measures the amount of agreement between the ground truth $\bm{G} =\{\bm{G}_1,\cdots ,\bm{G}_p\}$ and the clustering results $\bm{C} =\{\bm{C}_1,\cdots ,\bm{C}_p\}$. It is defined as:
	$$
	{\rm  SIM}(\bm{G},\bm{C})=\frac{1}{p}\sum\limits_{i=1}^p \mathop{{\rm max}}\limits_{1 \leq j \leq p}  {\rm Sim}(\bm{G}_i, \bm{C}_j),
	$$
	where ${\rm Sim}(\bm{G}_i, \bm{C}_j) = \frac{2|\bm{G}_i \cap \bm{C}_j|}{|\bm{G}_i| + |\bm{C}_j|}$, and $|\cdot|$ denotes the number of elements in the set. Similar to ARI, larger values of ${\rm SIM}$ indicate better clustering agreement with the ground truth. Unlike ARI, however, ${\rm SIM}$ is nonnegative and therefore cannot take values below $0$.

	\subsection{Competitive Approaches and Simulation Results}\label{res}
	For each experiment, we run 100 simulations and compute the average ARI and SIM values. In addition to the two proposed methods, we also include the following competing approaches:
	\begin{itemize}
		\item{ Ordinary Ward\textquotesingle s linkage (Ward): As a direct benchmark, we consider ordinary Ward's linkage adapted to functional observations. This method merges, at each step, the pair of clusters that produces the smallest increase in the within-cluster sum of squared deviations from the corresponding cluster mean functions. The deviations are measured using the norm of the underlying Hilbert space. For curves observed on a common grid, this norm is implemented through its discrete ($L^2$) approximation.}

		\item { Functional-median linkage (FM):
			For each cluster $\bm C$, we define the functional median as the
			curve having the largest MBD value among all curves in $\bm C$. The dissimilarity between two
			clusters is defined as the squared $L^2$ distance between their
			functional median curves.
		}
		
	\end{itemize}
	
	To further assess robustness, we include three methods that cluster
	individual curves rather than sets of curves. They are therefore treated
	as external pipeline-level benchmarks rather than direct linkage-level
	comparisons.
	\begin{itemize}
		\item TCLUST: A trimming-based robust clustering method
		\citep{fritz2012tclust}, implemented in the \texttt{tclust} package.
		It fits an ellipsoidal cluster model after trimming a proportion
		$\alpha$ of observations and constrains cluster scatter matrices to
		avoid degeneracy. We set $\alpha=1-\tau$.
		
		\item RSKC: A robust sparse $K$-means method implemented in
		\texttt{RSKC}. It combines trimming with sparse feature weighting,
		thereby performing robust clustering and variable selection. We set
		its trimming level to $\alpha=1-\tau$.
		
		\item OTRIMLE: A robust model-based method implemented in
		\texttt{otrimle}. It augments a Gaussian mixture with an improper
		uniform noise component, with parameters estimated by a
		(pseudo)-likelihood EM-type algorithm. We set the maximum
		noise/outlier proportion to $1-\tau$.
	\end{itemize}
	\begin{table}[!ht]
		\renewcommand{\arraystretch}{1.3} 
		\begin{center}
			\caption{Clustering results of Experiment 1 based on 100 simulations. The mean ARI and SIM values are reported.}
			\begin{tabular}{ccccccccccc}
				\toprule[1.2pt]
				Model  &Measure&$c$    &MS  &BD  &Ward  &    FM  &TCLUST  &RSKC  &OTRIMLE \\\midrule[1.2pt]
				&       & 0.1           &0.94            &\textbf{0.99}  &0.72    &0.94  &0.83  &0.86   &0.82\\ 
				Model 1 &SIM    & 0.15  &0.88            &\textbf{0.96}  &0.64    &0.86  &0.83  &0.86   &0.69\\
				&               & 0.2   &0.79            &\textbf{0.88}  &0.61    &0.77  &0.82  &0.85   &0.57\\\hline
				&       & 0.1           &\textbf{0.99}   &0.98           &0.77    &0.96  &0.83  &0.86   &0.82\\ 
				Model 2 &SIM     & 0.15 &\textbf{0.99}   &0.96           &0.68    &0.88  &0.83  &0.85   &0.68\\
				&       & 0.2           &\textbf{0.96}   &0.89           &0.63    &0.81  &0.83  &0.85   &0.57\\\hline
				&       & 0.1           &0.97            &\textbf{0.99}  &0.82    &0.79  &0.88  &0.90   &0.65\\ 
				Model 3 &SIM     & 0.15 &0.93            &\textbf{0.99}  &0.73    &0.76  &0.86  &0.87   &0.61\\
				&       & 0.2           &0.83            &\textbf{0.96}  &0.68    &0.74  &0.83  &0.84   &0.54\\\hline
				&       & 0.1           &\textbf{0.98}   &0.79           &0.75    &0.73  &0.72  &0.76  &0.59\\ 
				Model 4 &SIM     & 0.15 &\textbf{0.97}   &0.75           &0.69    &0.71  &0.72  &0.73  &0.59\\
				&       & 0.2           &\textbf{0.95}   &0.68           &0.66    &0.65  &0.71  &0.63 &0.58\\\midrule[1.2pt]

				&       & 0.1           &0.89    &\textbf{0.97}   &0.46    &0.88   &0.64    &0.70 &0.70\\ 
				Model 1 &ARI    & 0.15  &0.78    &\textbf{0.92}   &0.31    &0.75   &0.64    &0.69 &0.63\\
				&       & 0.2           &0.62    &\textbf{0.73}   &0.26    &0.58   &0.63    &0.70 &0.53\\\hline
				&       & 0.1           &\textbf{0.99}   &0.96    &0.56    &0.92   &0.65    &0.70 &0.69\\ 
				Model 2 &ARI    & 0.15  &\textbf{0.99}   &0.90    &0.38    &0.79   &0.65    &0.69 &0.62\\
				&       & 0.2           &\textbf{0.93}   &0.77    &0.29    &0.68   &0.65    &0.69 &0.51\\\hline
				&       & 0.1           &0.94   &\textbf{0.97}    &0.74    &0.64   &0.72    &0.76 &0.50\\ 
				Model 3 &ARI    & 0.15  &0.84   &\textbf{0.97}    &0.49    &0.61   &0.67    &0.69 &0.45\\
				&       & 0.2           &0.67   &\textbf{0.91}    &0.42    &0.56   &0.61    &0.63 &0.38\\\hline
				&       & 0.1           &\textbf{0.96}   &0.55    &0.52    &0.53   &0.49    &0.51 &0.47\\ 
				Model 4 &ARI    & 0.15  &\textbf{0.95}   &0.47    &0.40    &0.50   &0.49    &0.49 &0.41\\
				&       & 0.2           &\textbf{0.91}   &0.35    &0.34    &0.40   &0.47    &0.38 &0.35\\\midrule[1.2pt]
				
			\end{tabular}\\
			\label{tab1}
		\end{center}
	\end{table}

	\begin{table}
		\renewcommand{\arraystretch}{1.3} 
		\begin{center}
			\caption{Clustering results of Experiment 2 based on 100 simulations. The mean ARI and SIM values are reported.}
			\begin{tabular}{cccccccccc}
				\toprule[1.2pt]
				Model  &Measure& $c$   &   MS        & BD     &Ward       & FM  & TCLUST & RSKC & OTRIMLE \\\toprule[1.2pt]
				&               &0.1   &\textbf{0.98}       &0.83     &0.68   &0.73  & 0.64 & 0.67 & 0.66\\
				Eye-blink&SIM   &0.15  &\textbf{0.97}       &0.80     &0.62   &0.73  & 0.60 & 0.65 & 0.66\\
				&               &0.2   &\textbf{0.97}       &0.77     &0.59   &0.72  & 0.56 & 0.61 & 0.64\\\hline
				&               &0.1   &\textbf{0.95}       &0.81     &0.77   &0.73  & 0.40 & 0.76 & 0.66\\
				Eye-movement&SIM&0.15  &\textbf{0.95}       &0.80     &0.76   &0.73  & 0.30 & 0.64 & 0.65\\
				&               &0.2   &\textbf{0.93}       &0.80     &0.67   &0.73  & 0.20 & 0.58 & 0.64\\\midrule[1.2pt]
				
				&               &0.1   &\textbf{0.95}       &0.66     &0.38     &0.54   &0.65 & 0.46 & 0.55\\
				Eye-blink   &ARI&0.15  &\textbf{0.94}       &0.58     &0.28     &0.54   &0.59 & 0.36 & 0.51\\
				&               &0.2   &\textbf{0.94}       &0.54     &0.21     &0.54   &0.53 & 0.25 & 0.46\\\hline
				&               &0.1   &\textbf{0.92}       &0.60     &0.55     &0.55   &0.43 & 0.56 & 0.54\\
				Eye-movement&ARI&0.15  &\textbf{0.92}       &0.60     &0.55     &0.55   &0.37 & 0.41 & 0.48\\
				&               &0.2   &\textbf{0.88}       &0.58     &0.37     &0.56   &0.30 & 0.32 & 0.43\\\midrule[1.2pt]			  
			\end{tabular}\\
			\label{tab2}
		\end{center}
	\end{table}

	\begin{table}
		\renewcommand{\arraystretch}{1.3} 
		\begin{center}
			\caption{Average computation time (seconds).}
			\begin{tabular}{cccccccc}
				\toprule[1.2pt]
				$(N,T)$    &   MS        & BD            & Ward  & FM & TCLUST & RSKC & OTRIMLE \\\toprule[1.2pt]
				(600,100) &37.96  &  3.09 &1.30  &1.79   & 56.42     & 30.04 & 2.30\\
				(600,200)&69.75       & 5.48          &2.36  & 3.35     & 205.59    &186.56 & 2.77\\
				(1200,100)& 215.98       &29.40        &21.50  &24.71     & 74.56     & 62.36 & 2.84\\
				(1200,200) &306.01      & 60.39         &44.13  &50.02    & 277.79     & 175.33 & 7.19\\
				\midrule[1.2pt]
				
			\end{tabular}\\
			\label{taba}
		\end{center}
	\end{table}
	
	The results comparing the proposed linkages with the competing methods are presented in Tables~\ref{tab1} and~\ref{tab2}. In \textbf{Experiment~1}, when different types of contamination are introduced, especially as the contamination rate increases, the proposed linkages generally show strong robustness and achieve
	the best or competitive performance across most contamination settings. Since Model~1 involves purely magnitude contamination, BD-linkage demonstrates its effectiveness in handling this specific type of outlier. MS-linkage performs better on Model~2, which introduces shape contamination to the baseline model. In \textbf{Experiment~2}, our methods also show robustness against the two types of artifacts. The average computation times are reported in Table~\ref{taba} for different combinations of $N=30m$ and $T$, where $N$ denotes the total number of curves. { The computations were performed on a workstation equipped with an Intel Core i9-13900KF CPU (5.2 GHz) using a single CPU core.}
	
	The trimming-based competitors, TCLUST and RSKC, are relatively insensitive to the contamination rate $c$. One reason is that trimming removes a fixed fraction of observations that are least compatible with the current fitted cluster model at each iteration, which is effective when contaminated curves are globally atypical. However, in our simulations, some contaminated observations from one cluster become highly similar to the majority pattern of another cluster. Such observations are not necessarily far from the fitted model and therefore are unlikely to be trimmed; instead, they may be absorbed as normal curves in the competing cluster, thereby reducing clustering accuracy.
	
\begin{figure}[htbp]
		\centering
		\includegraphics[width=0.8\textwidth]{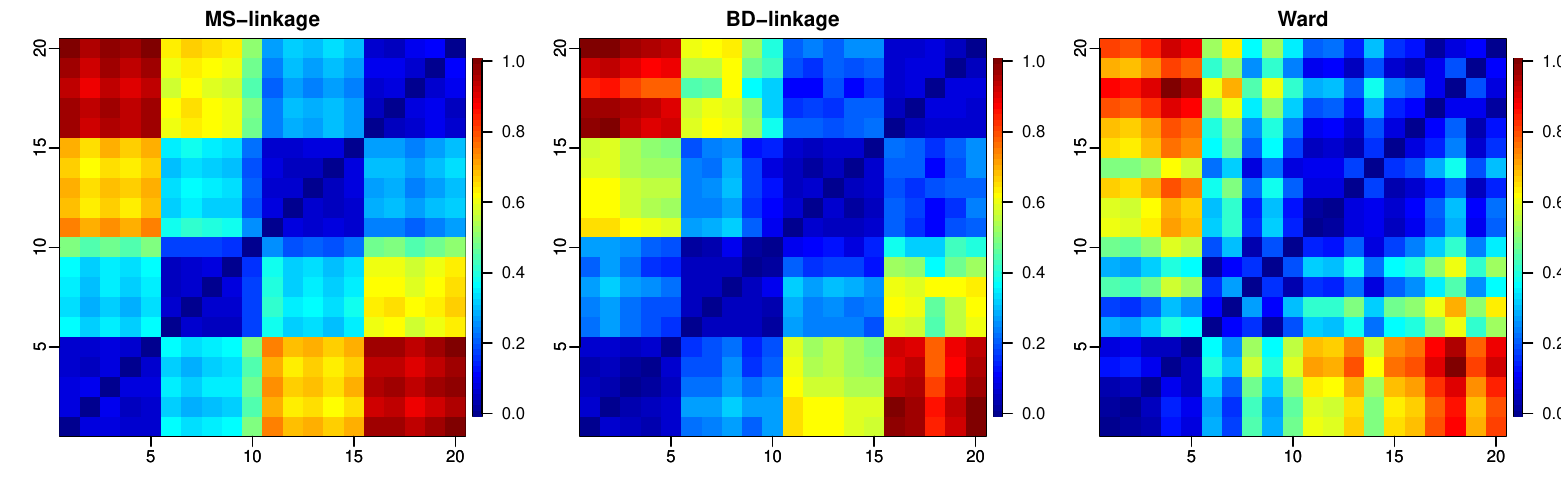}
		\caption{Linkage matrices of the initial clusters in Experiment~1 under Model~2 with $c=0.1$.}
		\label{dis}
	\end{figure}

	Figure~\ref{dis} further illustrates the robustness of the proposed criteria by presenting the linkage matrices of the initial clusters in \textbf{Experiment~1} (Model~2 with $c=0.1$). The linkage matrices of the two proposed methods exhibit much clearer block structures than that of ordinary Ward\textquotesingle s linkage.
	
	\subsection{Initial-Cluster Size and Tuning Parameter Selection}
	\label{sen}
\begin{figure}[htbp]
		\centering
		\subfigure[]{\includegraphics[width=0.4\textwidth]{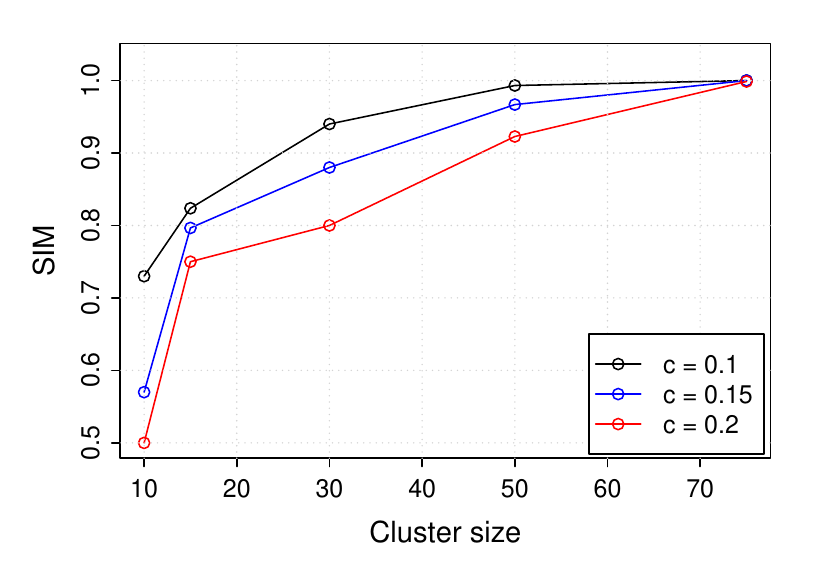}}
		\subfigure[]{\includegraphics[width=0.4\textwidth]{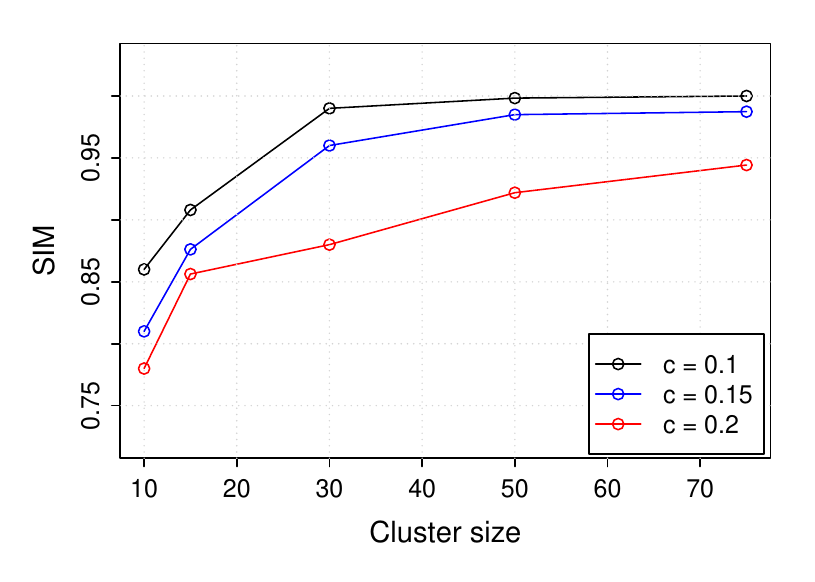}} 
		\subfigure[]{\includegraphics[width=0.4\textwidth]{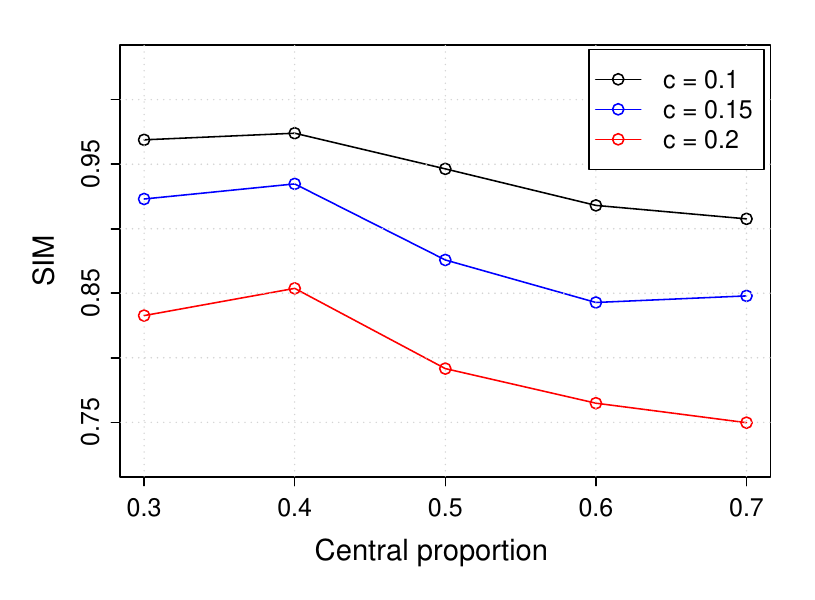}}
		\subfigure[]{\includegraphics[width=0.4\textwidth]{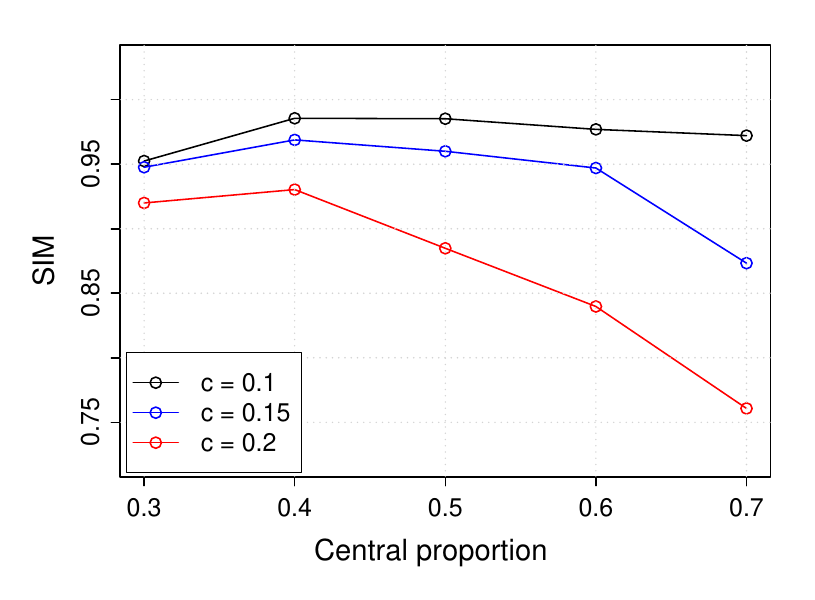}} 
		\caption{Top panel: the clustering accuracy versus the initial cluster size. (a) MS-linkage and (b) BD-linkage.
			Bottom panel: clustering accuracy versus the central proportion
			$\tau$. (c) MS-linkage and (d) BD-linkage.}
		\label{ratio}
	\end{figure}

	{We use Model~1 of \textbf{Experiment~1} to investigate the effects of the
		initial-cluster size and the central proportion $\tau$. We first examine
		the initial-cluster size while fixing $\tau=0.5$ and keeping all other
		simulation settings unchanged. Each true cluster contains 150 curves and
		is partitioned into initial clusters of equal size. We consider
		\[
		m_0\in\{10,15,30,50,75\},
		\]
		where $m_0=30$ is the default setting used in the main simulation. In particular, $m_0=10$ falls below the practical threshold of the
		standard MS construction and therefore invokes the small-cluster rule
		described in Section~\ref{initial_size}.
		These values are selected because they are divisors of 150, ensuring that
		each true cluster can be divided into initial clusters of equal size. The
		corresponding numbers of initial clusters within each true cluster are
		$15$, $10$, $5$, $3$, and $2$, respectively, yielding $60$, $40$, $20$,
		$12$, and $8$ initial clusters in total. For each value of $m_0$, we
		conduct 100 Monte Carlo replications and report the average SIM values in
		Figure~\ref{ratio}(a) and (b).
		
		For both linkages, clustering accuracy generally improves as the
		initial-cluster size increases, indicating that larger initial clusters
		provide more stable centrality rankings and central-band estimates. The
		deterioration becomes increasingly pronounced as the initial-cluster size
		is reduced, particularly when $m_0$ decreases from 15 to 10. This effect
		is more evident for the MS-linkage than for the BD-linkage, suggesting
		that the MS-based central-subset construction is more sensitive to very
		small initial clusters. In contrast, the improvement becomes much more
		modest once the initial-cluster size exceeds 30. We therefore use
		$m_0=30$ throughout the main simulation studies as a practical compromise
		between clustering accuracy, stability, and computational cost. In real
		applications, however, sufficiently large initial clusters may not always
		be available, for example when a time series is too short to generate
		many segments. We therefore recommend using initial clusters that are as
		large as the natural data structure permits.

		 We also examine how the central-curve proportion $\tau$ affects the 
		clustering accuracy. By setting
		$\tau=0.3,0.4,0.5,0.6$, and $0.7$, we obtain Figure~\ref{ratio} (c) and (d), which report the
		SIM values for the MS- and BD-linkages under different contamination
		rates. For both linkages, clustering accuracy generally increases slightly
		from $\tau=0.3$ to $\tau=0.4$ and then decreases as $\tau$ becomes
		larger. The best or nearly best performance is generally attained
		around $\tau=0.4$. When $\tau$ is too small, the central subset may
		contain insufficient information to distinguish the clusters
		effectively. In contrast, a large value of $\tau$ increases the
		probability that contaminated curves are included in the central
		subset, thereby reducing robustness. This deterioration becomes more
		pronounced as the contamination rate $c$ increases, particularly for
		larger values of $\tau$. These results illustrate the trade-off between
		retaining sufficient representative information and limiting the
		influence of contaminated curves. Although $\tau=0.4$ yields the highest SIM values in most of these
		settings, $\tau=0.5$ remains close to the optimal region and provides a
		common, prespecified choice across all simulation models.}
	
	\subsection{Sensitivity to Initial-Cluster Impurity}
	
\begin{figure}[htbp]
		\centering
		\subfigure[]{\includegraphics[width=0.4\textwidth]{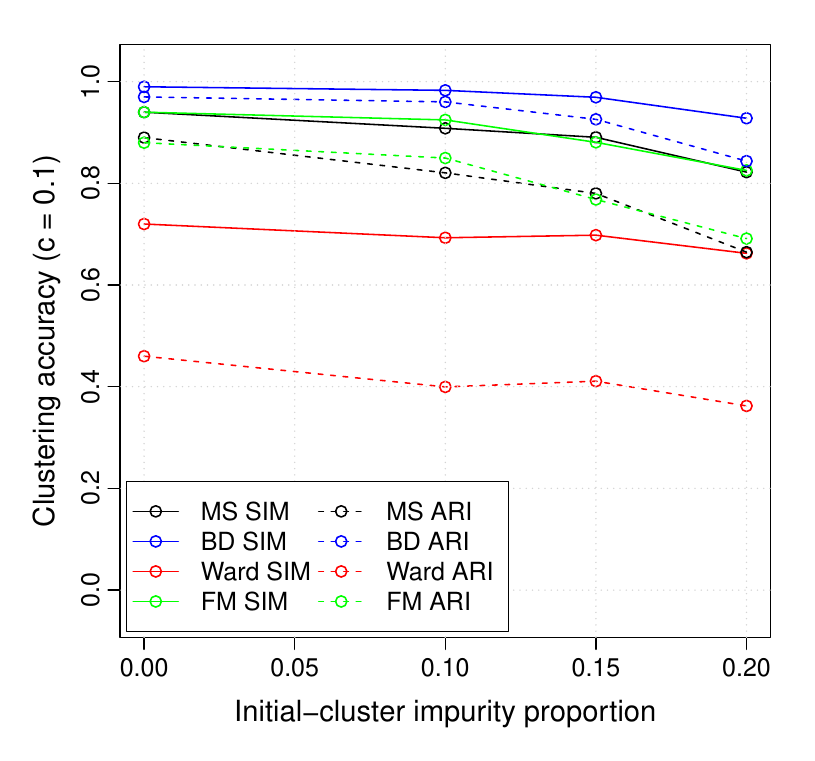}}
		\subfigure[]{\includegraphics[width=0.4\textwidth]{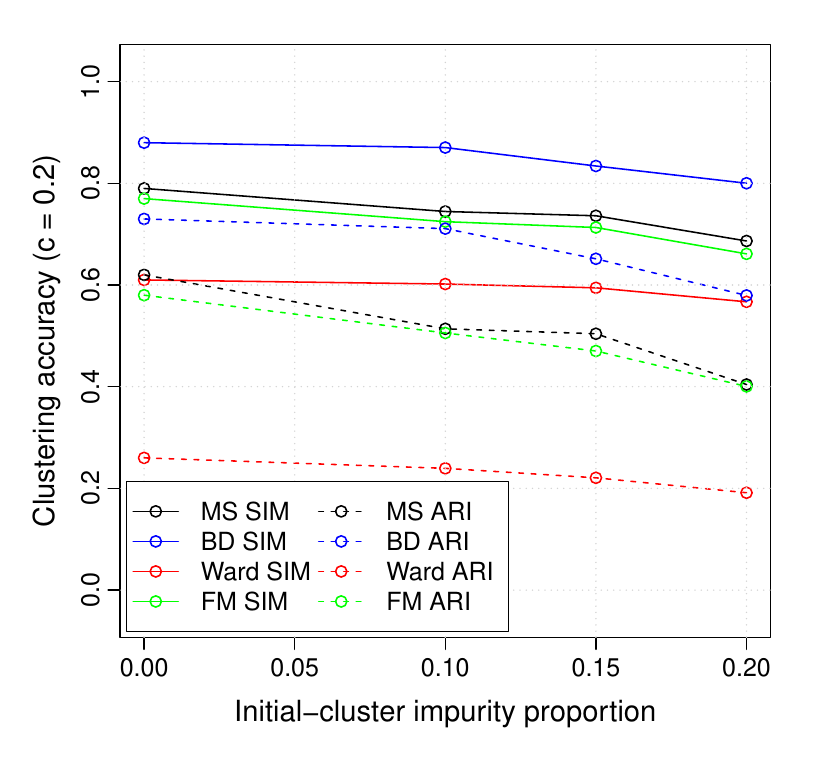}}
		\caption{Clustering accuracy as a function of the initial-cluster
			impurity proportion $\eta$. Panels (a) and (b) correspond to
			$c=0.1$ and $c=0.2$, respectively.}
		\label{imp}
	\end{figure}

	{
		The preceding simulations construct each initial cluster predominantly
		from curves belonging to the same latent class. To investigate the
		sensitivity of the set-level clustering methods to imperfect initial
		grouping, we introduce an initial-cluster impurity proportion
		\[
		\eta\in\{0,0.10,0.15,0.20\}.
		\]
		For an initial cluster $\bm C_j$ containing $n_j$ curves, we randomly
		select
		\[
		q_j(\eta)=\left\lfloor \eta n_j\right\rfloor
		\]
		of its members and exchange them with the same number of curves from
		initial clusters associated with different latent classes. The exchange
		is performed so that the size of each initial cluster remains unchanged.
		Consequently, approximately a proportion $\eta$ of the curves in each
		initial cluster originates from other true classes, while the remaining
		curves preserve the original class structure.
		
		This impurity mechanism is conceptually different from the outlier
		contamination considered above. The contamination proportion $c$
		controls the proportion of atypical curves generated through magnitude,
		shape, covariance, or artifact contamination, whereas $\eta$ controls
		the degree of mixing among different latent classes within the initial
		clusters. Thus, this experiment evaluates the sensitivity of the
		proposed procedure to the quality of the initial grouping rather than
		its robustness to curve-level outliers.
		
		For each Monte Carlo replication and each value of $\eta$, the same
		impure initial partition is supplied to all set-level hierarchical
		methods. Since TCLUST, RSKC, and OTRIMLE operate on individual curves
		and do not use the initial grouping, their results are unaffected by
		$\eta$ and are therefore excluded from this sensitivity analysis.
		We illustrate the effect of initial-cluster impurity using Model~1 with
		$c=0.1$ and $c=0.2$, representing relatively low and high contamination
		levels, respectively. The corresponding results are shown in
		Figure~\ref{imp}. As shown, the performance of the proposed MS- and
		BD-linkages generally decreases as the initial-cluster impurity
		proportion $\eta$ increases, indicating that the methods are affected
		to some extent by the quality of the initial grouping. However, the
		deterioration is relatively moderate, particularly for
		$\eta\leq 0.15$, and both proposed linkages retain comparatively high
		ARI and SIM values under both contamination levels. In contrast, Ward
		and FM show smaller absolute decreases in some cases, but their baseline
		clustering accuracies are already substantially lower, so the smaller
		decline should not be interpreted as greater overall robustness. These
		results suggest that, although the proposed methods benefit from
		reasonably homogeneous initial clusters, their usefulness is not
		restricted to nearly pure initial groupings and they remain effective
		under a moderate degree of initial-cluster impurity.
	}
	
	\section{EEG Data Analysis}\label{APP}
	In this section, we apply the proposed linkages to resting-state
	multichannel EEG data and single-channel epilepsy EEG recordings.
	Both analyses are conducted in the frequency domain using the SDF
	estimator described in Section~\ref{SM}. Following Section~\ref{tau},
	we set $\tau=0.7$ for both applications.
	
	\subsection{Resting-state EEG Data Analysis}\label{eegd}
\begin{figure}[htbp]
		\begin{center}
			\subfigure[]{\includegraphics[width=0.5\textwidth]{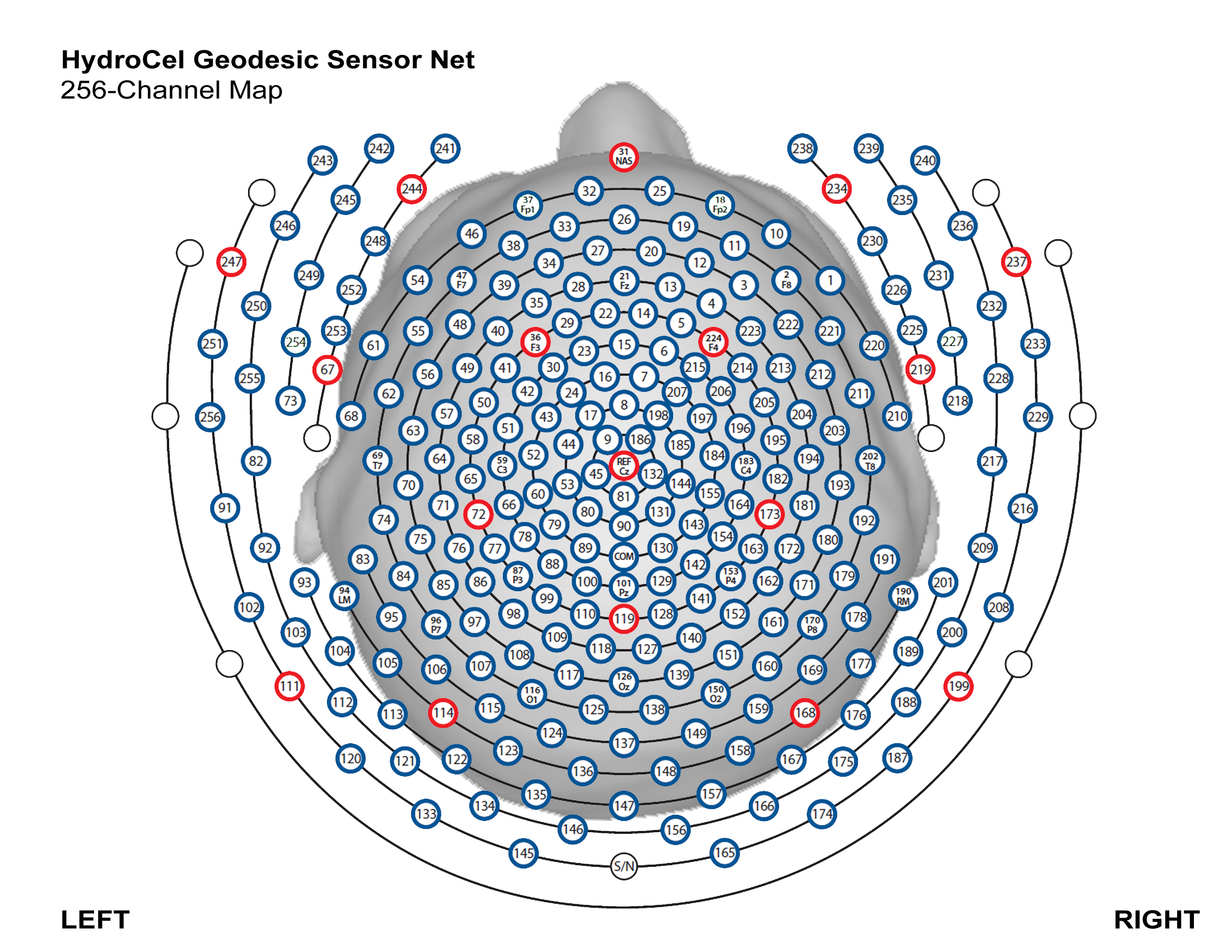}}\\
			\subfigure[]{\includegraphics[width=0.4\textwidth]{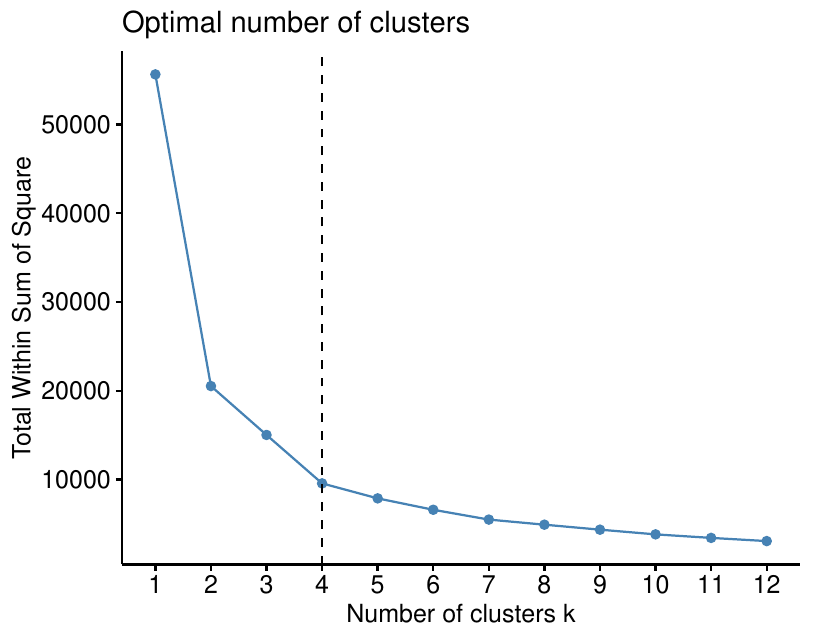}}
			\subfigure[]{\includegraphics[width=0.4\textwidth]{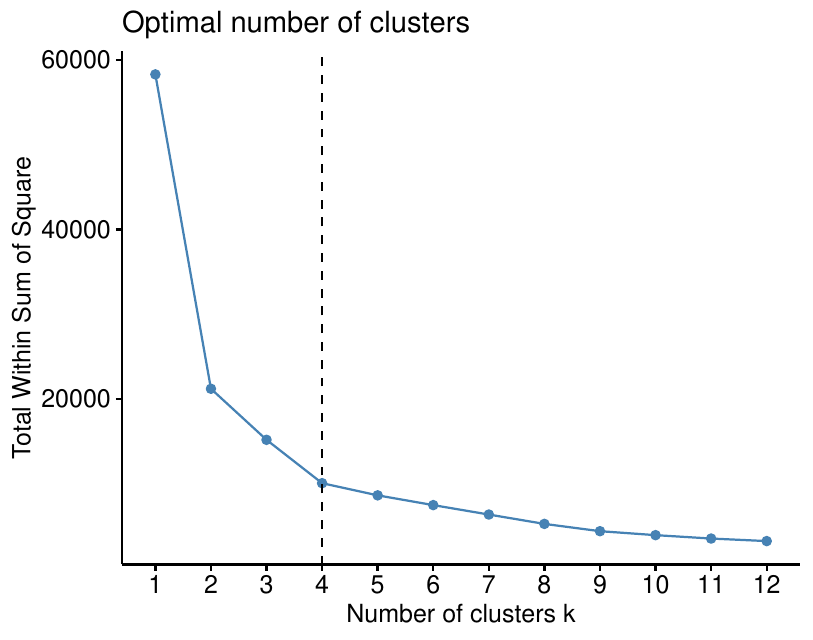}}
		\end{center}
		\caption{(a) Locations of the $256$ channels on the scalp surface. (b) The scree plot of the first minute. (c) The scree plot of the last minute.}
		\label{eeg}
	\end{figure}

	The first application uses resting-state EEG data from Wu et al.~\citeyearpar{wu2014resting}. The goal is to cluster EEG channels with similar spectral characteristics.  The EEG signals were recorded from 256 scalp channels (Figure~\ref{eeg}(a)) at 1000~Hz. Among the 256 channels, 62 were excluded due to data-quality issues, leaving 194 channels for analysis. The total recording duration is three minutes, and we analyze the first minute and the last minute separately. Within each minute, we partition the data from each channel into 60 non-overlapping 1-second segments and estimate an SDF for each segment. Thus, for a given minute, each channel is represented by a set of 60 estimated SDFs, which we treat as the initial cluster for that channel. We then apply the proposed hierarchical clustering to the 194 initial clusters to obtain a clustering of channels for that minute. The optimal number of clusters $p$ is selected by the elbow method implemented in the \texttt{R} package \texttt{factoextra}. We plot the total within-cluster sum of squares against the number of clusters (ranging from 1 to 12). As illustrated in Figure~\ref{eeg}(b,c), the ``elbow'' appears at 4 clusters for both the first and the last minutes.
	
\begin{figure}[htbp]
		\begin{center}
			\subfigure{\includegraphics[width=0.65\textwidth]{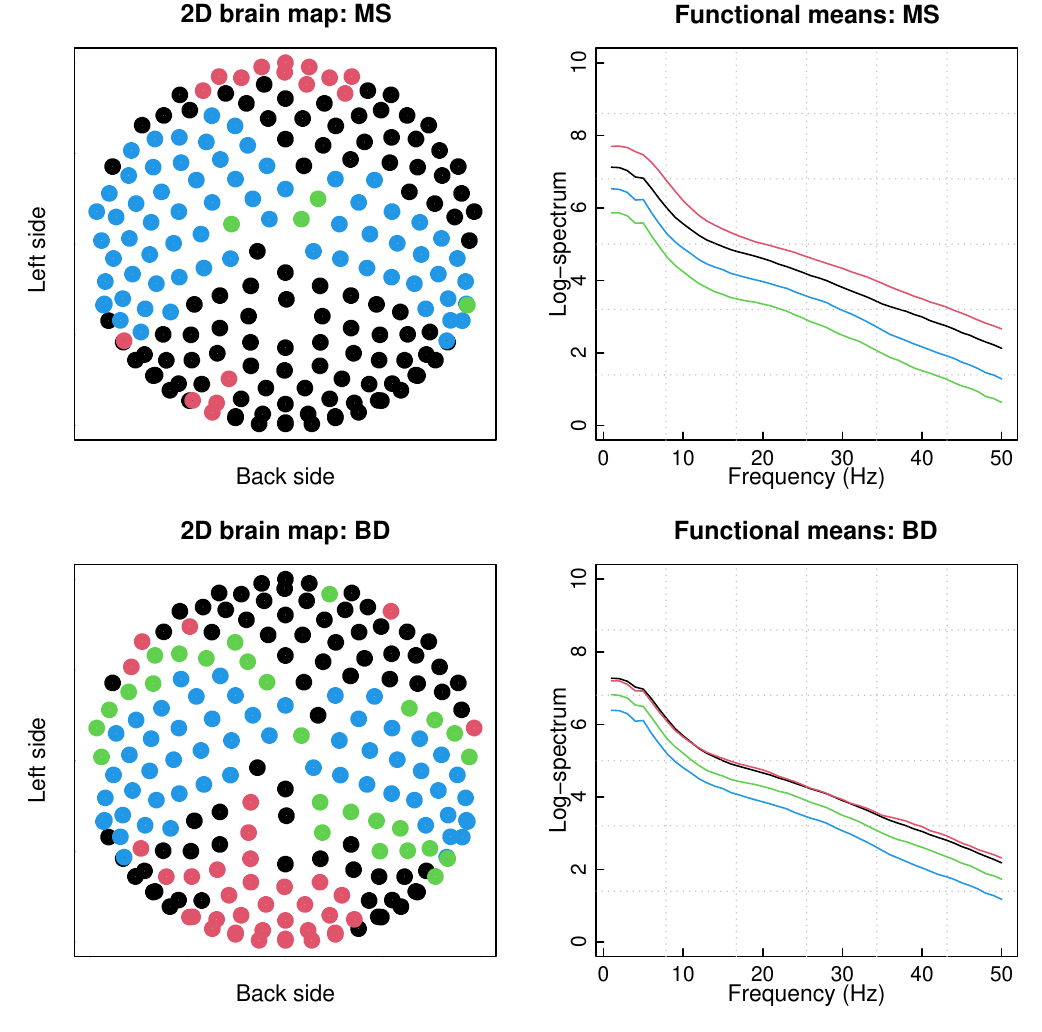}}\\
		\end{center}
		\caption{The clustering results of the first minute. Top row: MS-linkage; bottom row: BD-linkage}
		\label{first}
	\end{figure}

\begin{figure}[htbp]
		\begin{center}
			\subfigure{\includegraphics[width=0.65\textwidth]{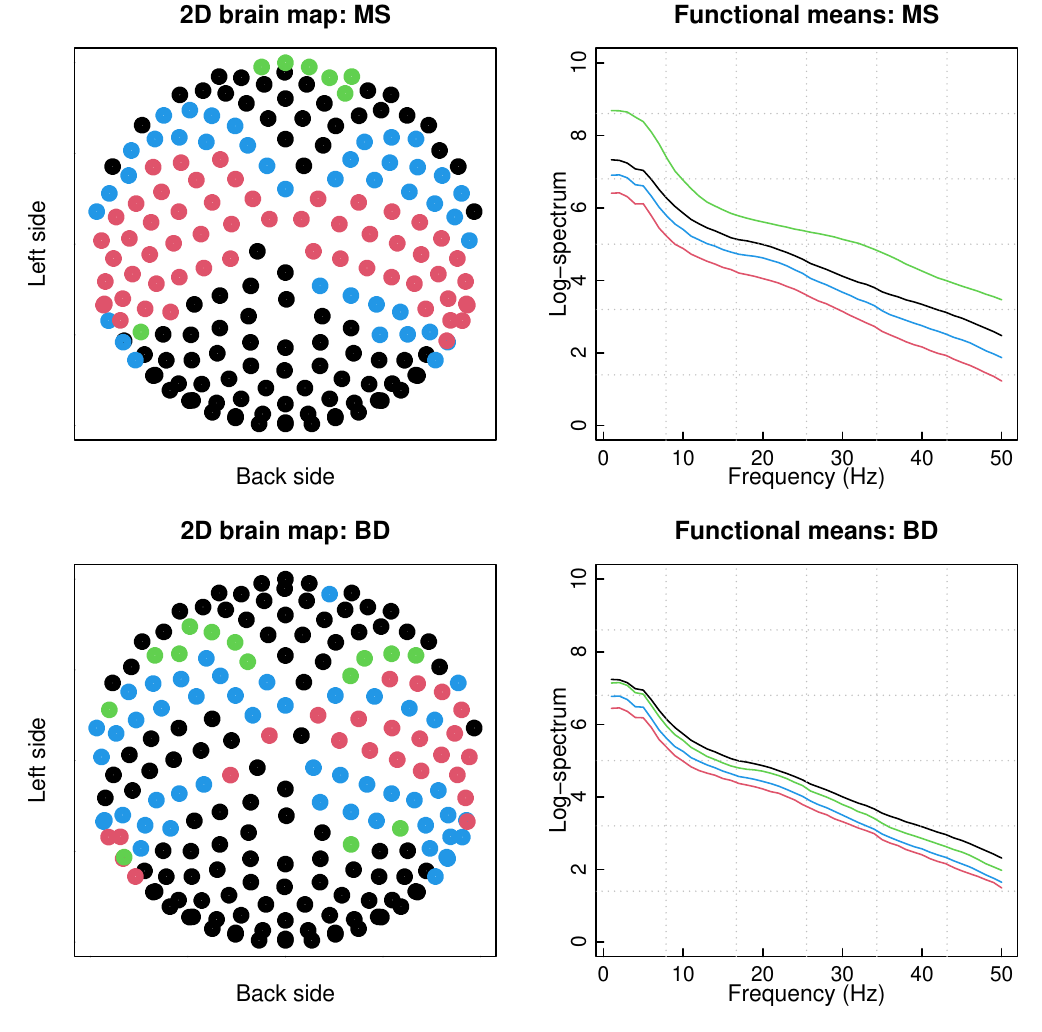}}\\
		\end{center}
		\caption{The clustering results of the last minute. Top row: MS-linkage; bottom row: BD-linkage}
		\label{last}
	\end{figure}

	We illustrate the clustering results using 2D brain maps and the functional means of each cluster. Figure \ref{first} presents the clustering results of the EEG records from the first minute. Out of the 194 channels,  MS-linkage assigns 107, 67, 16, and 4 channels to the four clusters, respectively, while BD-linkage assigns 73, 57, 37, and 27 channels. For the last minute, as illustrated in Figure \ref{last}, MS-linkage assigns 89, 57, 41, and 7 channels to the four clusters, respectively, while BD-linkage assigns 108, 48, 24, and 14 channels. 
	
	As shown in Figures \ref{first} and \ref{last},  
	\begin{itemize}
		\item For both stages (first and last minutes) and both linkages, the elbow method suggests using $p=4$ clusters.
		\item All brain maps show relatively well-separated, spatially homogeneous, and approximately left--right symmetric regions.
		\item According to the brain maps and the corresponding functional means, the channels in the middle region of the brain exhibit lower density at 0\textendash50 Hz than the front and back regions.

		\item { The results from both the first and last minutes exhibit the three
			general characteristics described above. The consistency of these broad spatial and spectral patterns is
			reasonable because the data were recorded from the same subject under
			the same resting-state condition.}
	\end{itemize}
	
	
	Clustering channels by spectral density provides a compact summary of
	spectral similarity and spatial organization. Mapping the clusters onto
	the scalp facilitates interpretation of left--right symmetry and
	frequency-specific power patterns, while the resulting channel groups
	can support downstream connectivity or network analyses
	\citep{euan2018hierarchical,maadooliat2018nonparametric,Li03072019,
		stam2007phase,bressler2010large}.

	\subsection{Epilepsy EEG Data Analysis}\label{eard}
	{  In this application, we use the EEG data described in
		Andrzejak et al. \citeyearpar{andrzejak2001indications} to investigate the
		clustering of EEG recordings from different physiological and
		pathological states. There are five sets of data (denoted as A, B, C, D and E) collected from individuals in different states. Each set contains 100 single-channel EEG segments of length 4097 in 23.6 seconds. To construct the initial clusters without using the known A--E group
		labels, we treat each original EEG time series as an independent
		observational unit. Specifically, for each time series, we use the first
		4000 observations and divide them into 20 nonoverlapping segments of
		length 200. A spectral density curve is estimated from each segment,
		yielding 20 spectral curves from each original EEG time series. These
		20 curves are grouped together to form one initial cluster. Thus, each initial cluster is determined solely by the identity of the
		original EEG recording and does not use information about whether the
		recording belongs to set A, B, C, D, or E. The A--E labels are used to
		define the benchmark clustering tasks and, after clustering, to
		evaluate the resulting partitions, but they are not used in the
		construction of the initial clusters or in the agglomerative clustering
		procedure.}
	
	The content of the five sets is:
	\begin{itemize}
		\item A: Healthy volunteers with eyes open;
		\item B: Healthy volunteers with eyes closed;
		\item C: Epilepsy patients in seizure-free intervals from the
		hemisphere opposite the epileptogenic zone;
		\item D: Epilepsy patients in seizure-free intervals from within
		the epileptogenic zone;
		\item E: Epilepsy patients in seizure intervals.
	\end{itemize}  
	
	Following Maadooliat et al. \citeyearpar{maadooliat2018nonparametric}, we
	consider four tasks of increasing difficulty:
	\begin{itemize}
		\item Easy (A vs. E): 200 initial clusters are grouped into $p=2$ clusters.
		\item Intermediate (D vs. E): 200 initial clusters are grouped into $p=2$ clusters.
		\item Hard (A vs. D): 200 initial clusters are grouped into $p=2$ clusters.
		\item Challenging (A vs. D vs. E): 300 initial clusters are grouped into $p=3$ clusters.
	\end{itemize}

	We also compare our methods with two alternative approaches, NCSDE and GLK, as used in Maadooliat et al. \citeyearpar{maadooliat2018nonparametric}, using ARI and SIM as evaluation metrics. Specifically, NCSDE is a collective spectral density estimation and clustering method for multiple time series based on a Whittle-likelihood criterion. It employs a low-dimensional basis representation to capture similarities among the log-spectral densities and achieve dimension reduction, and the resulting basis coefficients are then used for clustering. In GLK, the dissimilarity between two time series is computed using an adaptation of the generalized likelihood ratio test for assessing the equality of two log-spectra.
	
	The results are presented in Table~\ref{tab3}, which summarizes the
	clustering performance on the epilepsy EEG data. { In the Easy task
		(A vs.~E), the proposed MS-linkage achieves perfect clustering, with
		both ARI and SIM equal to 1, while the BD-linkage also performs well,
		with ARI 0.97 and SIM 0.87. NCSDE gives similarly high accuracy in this
		setting. For the Intermediate task (D vs.~E), the three methods perform
		very similarly: MS and BD both obtain an ARI of 0.87 and a SIM of 0.97,
		while NCSDE achieves an ARI of 0.88 and the same SIM value of 0.97.
		
		The differences become more pronounced in the more difficult tasks. For
		the Hard task (A vs.~D), MS and BD achieve ARI values of 0.72 and 0.63,
		respectively, with corresponding SIM values of 0.93 and 0.90, whereas
		NCSDE attains higher values of 0.85 and 0.96. For the Challenging task
		(A vs.~D vs.~E), MS achieves an ARI of 0.76 and a SIM of 0.92, while BD
		obtains an ARI of 0.61 and a SIM of 0.86. NCSDE again gives the highest
		accuracy, with an ARI of 0.85 and a SIM of 0.95. Thus, the proposed
		linkages are highly competitive in the Easy and Intermediate settings.
		In the more difficult tasks, NCSDE performs better overall, although
		MS-linkage still achieves relatively high SIM values, particularly in
		the Challenging task.
	}
	
	The stronger performance of NCSDE in the Hard and Challenging tasks may
	be partly attributed to its methodological design. NCSDE is specifically
	developed for the joint estimation and clustering of multiple groups of
	time series, allowing information to be shared across series during
	spectral estimation. Such collective estimation can be advantageous when
	the spectral differences between groups are relatively small and the
	estimation noise is non-negligible. In contrast, our procedure estimates
	the spectrum of each segment separately and subsequently introduces
	robustness through the MS- or BD-based central-subset linkage.
	Overall, these results indicate that the proposed methods remain
	effective for clustering grouped EEG spectral curves, although their
	advantage is not uniform across all clustering tasks.
	
	Clustering epilepsy EEG recordings provides an unsupervised
	grouping of recordings with similar spectral characteristics, which may reflect latent heterogeneity relevant for further clinical investigation. Moreover, the resulting cluster assignments can serve as weak (pseudo-)labels to facilitate downstream supervised tasks, such as seizure detection or seizure/non-seizure classification.
	\begin{table}[ht]
		\renewcommand{\arraystretch}{1.3} 
		\begin{center}
			\caption{ Clustering results of the four methods. The ARI and SIM are reported.}
			\begin{tabular}{cccccc}
				\toprule[1.2pt]
				Task & Measures       &   MS  & BD        & NCSDE  & GLK   \\\toprule[1.2pt]
				Easy & ARI            & {\bf 1}  &  0.97  &0.98 &  0.36  \\
				A vs. E	&  SIM   &        {\bf 1}  &0.87      &0.99 &  0.80\\\hline
				Intermediate & ARI  &  0.87  & 0.87     &\textbf{0.88} &  0.28 \\
				(D vs. E)& SIM       & \textbf{0.97}  &\textbf{ 0.97 }     &\textbf{0.97} & 0.75\\\hline
				Hard & ARI    &  0.72  &0.63   &\textbf{0.85} &  0.43 \\
				(A vs. D)& SIM &         0.93    &0.90   &\textbf{0.96} &0.83\\\hline	
				Challenging & ARI    &  0.76     &0.61 &\textbf{0.85} &  0.25 \\
				(A vs. D vs. E)& SIM   & 0.92     &0.86  &\textbf{0.95} &0.66\\\toprule[1.2pt]
				
			\end{tabular}\\

			\label{tab3}
		\end{center}
	\end{table}
	
	\FloatBarrier
	\section{Conclusion}\label{CO}
	In this paper, we proposed two robust nonnegative functional Ward-type linkage criteria. For each candidate pair of clusters, central curves are selected separately within the two clusters using magnitude-shape outlyingness or modified band depth, and the linkage is defined as the weighted increase in band width required to form a common band containing both central subsets. The proposed criteria thereby reduce the influence of outliers and contamination on the agglomerative merging procedure. Their robustness is demonstrated through simulations involving different types of outliers in the temporal and spectral domains and through applications to two EEG datasets.
	However, the proposed linkages are computationally more demanding than
	ordinary Ward's linkage, particularly the MS-linkage, because of the
	repeated centrality evaluation and two-dimensional contour estimation.
	For very large functional datasets, computational efficiency therefore
	remains an important practical consideration. Possible simplifications
	include reducing the number of boundary segments of the estimated
	contour or approximating it with a simpler polygon, such as a
	rectangle. A penalty on the number of contour arcs may also be
	introduced to regulate its geometric complexity. These modifications
	may reduce computational cost, but they may also alter the selected
	central subset and affect clustering accuracy. Investigating this
	trade-off provides a useful direction for future work.

	MS-linkage selects central curves via the MS-plot, where directional outlyingness is summarized by $\mathrm{MO}$ and $\mathrm{VO}$; consequently, it tends to exclude curves with global shifts or localized deformations. BD-linkage instead ranks curves by $\mathrm{MBD}$, a depth-based notion of centrality that reflects how often a curve lies within empirical bands. In our simulations (Experiment~1, Section~3), BD-linkage performs better under Type~1 contamination dominated by magnitude shifts, whereas MS-linkage is more effective under Type~2 contamination involving both magnitude and shape changes. In the simulation settings considered here, MS-linkage tends to be
	more effective when shape contamination is present, whereas BD-linkage
	performs particularly well under primarily magnitude-driven
	contamination. MS-linkage may provide a useful general-purpose choice when the nature
	of the contamination is unknown. On the other hand, if (i) a depth-based notion of centrality is desired, (ii) the main concern is to downweight curves with consistently low depth, or (iii) computational efficiency is a primary concern, then BD-linkage is preferable. 
	
	In closing, further developments are needed. First, it is worthwhile to extend the proposed methods to multivariate functional data and to explore additional approaches for selecting the most central curves. Second, modeling a wider range of EEG artifacts can help further validate the robustness of the proposed linkages. Third, future work will focus on reducing the computational complexity of the proposed methods.

	\section*{Acknowledgments}
	The authors thank Professor Jennifer Wu, Ralph G. Andrzejak and Mehdi Maadooliat for sharing the EEG data sets. We also thank the two reviewers for their valuable comments and suggestions, which have substantially improved the quality of this paper.
	
	\section*{Funding}
	Chen's research was partially supported by the National Natural Science Foundation of China (No. 12301326), University Natural Science Research Project of Anhui Province (No. 2023AH050099), and Natural Science Foundation of Anhui Province (No. 2308085QA05). Zhu's research was partially supported by the National
	Natural Science Foundation of China (No. 12201218) and the University Natural Science Research Project of Anhui Province (No. 2023AH033436).
	
	\section*{Data availability}
	The \texttt{R} code to reproduce the results, including all functions used in this paper, is available at \url{https://github.com/tianbochen1/Functional-Ward-s-Linkages}. We also uploaded the original epilepsy EEG data, as well as their estimated SDFs to the repository.
	
	\section*{Conflict of interest}
	
	The authors declare no conflicts of interest.

	\bibliographystyle{apalike}
	\bibliography{reff}

\end{document}